# Additional Boundary Condition for the Wire Medium

Mário G. Silveirinha[1]

Universidade de Coimbra[1], Instituto de Telecomunicações[1]
Departamento de Eng. Electrotécnica, Pólo II, 3030 Coimbra, Portugal, mario.silveirinha@co.it.pt

*Abstract* – **In this paper, it is proved that the continuity of the tangential components of the average electric and magnetic fields is insufficient to describe the reflection of plane waves by a set of thin parallel wires embedded in a dielectric host using a homogenization approach. It is shown that an additional boundary condition is required to conveniently model a slab of the homogenized metamaterial. In order to understand how the different electromagnetic modes are excited at the interface, the problem of reflection of a plane wave by a set of semi-infinite parallel wires is solved analytically within the thin-wire approximation. Based on the derived result and other arguments, a new boundary condition is proposed for the homogenized wire medium. Extensive numerical simulations support our theoretical analysis, and show that when the additional boundary condition is considered the agreement between full wave results and homogenization theory is very good even for wavelengths comparable with the lattice constant.**

*Index Terms* – metamaterials, homogenization theory, wire medium, spatial dispersion, boundary conditions, ABC.

## I. INTRODUCTION

The characterization of the wire medium is an important problem in electromagnetics. The reason is that this metamaterial is suitable for fabrication at the microwave band and can be operated in a regime where it effectively behaves as a negative permittivity material [1]-[2].

In recent years, there has been a great interest in materials with negative (or near zero) permittivity. For example, it was shown that if not only the permittivity but also the permeability of the medium is negative (DNG medium) then negative refraction occurs [3]. In [4] it was shown that a source embedded in a metamaterial with permittivity near zero radiates most of the energy into a narrow angular cone. In [5] it was shown that an $\varepsilon$-negative material can be tailored to synthesize a dielectric crystal that behaves effectively as a DNG medium. In [6] it was proved that $\varepsilon$-negative (and near zero) materials can be used as covers to significantly reduce the scattering from moderately large particles.

Because of these and other prospective applications of $\varepsilon$-negative materials, it is important to rigorously characterize the wire medium. Although the first studies about this composite structure were published more than four decades ago [1] there are still some open research topics, and recent works show that the propagation in the wire medium is not as simple as it was initially thought. It was recently proved that all the common geometries of the wire medium are characterized (at least to some extent) by spatial dispersion [7]-[8], and that consequently the constitutive relations in the homogenized structure are non-local [9]. Also, in [10] it was shown that in a 2D-wire mesh of non-connected wires the average fields at an interface cannot be identified with those of the bulk metamaterial.

As is well-known, when spatial dispersion occurs the usual boundary conditions (i.e. the continuity of the tangential components of **E** and **H**) are insufficient to describe the field interaction at an interface. In fact, additional boundary conditions (ABCs) are required [11]-[17]. The ABC concept has been used in electromagnetics of spatial dispersive media and in solid-state physics for many decades [11]. Unfortunately, no general theory is available to derive such conditions, which in general are dependent of the local geometry/properties of the medium. Some classical ABCs, which have been used in a wide range of problems, were formulated in [15] and impose that either the polarization vector or the derivative of the polarization vector are zero at the interface. These and other ABCs are not suitable (at least *a priori*) to the problem we want to study here, since as referred before, the spatial dispersion phenomenon is specific to the microstructure of the medium.

In this paper, we propose a new boundary condition to characterize the (homogenized) wire medium-1D (array of parallel thin metallic wires). The only work that we could find in the literature directly related with this topic is the "ABC-free" [16] motivated approach reported in [17]. In that work, it is shown that the classical ABCs are not appropriate to homogenize the wire medium problem. Instead, the authors of [17] suggest introducing a transition layer between the air region and the array of wires. A source that is excited by the incident wave is placed at the transition layer, and determines indirectly the amplitude of the scattered fields. This approach is interesting, but the choice of source in [17] seems to be somehow arbitrary and the proposed model was not numerically tested.

Our approach here is very different. Based on physical arguments we propose a new boundary condition to characterize the interaction of the fields in free-space and the fields in the homogenized wire medium. The proposed boundary condition is successfully tested against full wave numerical data. To further demonstrate the generality of the ABC, we theoretically calculate the reflection coefficient of a plane wave impinging on a set of semi-infinite parallel wires. An exact analytical formula for the reflection coefficient is derived under the thin wire assumption, using complex function theory and a generalization of the method proposed in [18]-[20], and it is shown that the result is completely consistent with the new ABC. It is important to note that even though other studies [18]-[20] have shown that in some cases it is possible to solve exactly the reflection problem in a semi-infinite crystal, all these works assumed that the inclusions could be replaced by point-dipole oscillators. However, such assumption cannot be used in this paper because here the inclusions are infinitely long metallic wires.

The organization of the paper is as follows. In section II, the new boundary condition is proposed. In section III, it is shown that the results predicted by the analytical model compare very well (even for moderately small wavelengths) with full wave numerical simulations. To further support the new ABC in sections IV and V, we study the problem of reflection of a plane wave by an array of semi-infinite wires. Firstly, in section IV, we briefly review the necessary formalism to tackle the reflection problem. In [18]-[20] it was shown that it is possible (for the case of point dipole oscillators) to relate the amplitudes of the excited modes in the semi-infinite crystal with the electric dipole moment and the wave vector of the modes. In [20] it was proved that these results are actually valid for an arbitrary dielectric crystal with no restrictions on the electrical size of the inclusions. In this paper we present an extension of these results to the metallic case. In section V, we apply the derived formalism to calculate the reflection coefficient analytically. Following an idea proposed in [20], we will show that the solution of the problem is intrinsically related with the reconstruction of an analytical function from the knowledge of its zeros and poles. Although the idea proposed in [20] is relatively plain, things are not really so simple because the convergence of some infinite products and the uniqueness of the solution may be questionable. We present a rigorous analysis that solves these issues unambiguously and we show that the result is compatible with the new ABC. Finally, in section VI, the conclusions are presented.

## II. ADDITIONAL BOUNDARY CONDITION FOR THE WIRE MEDIUM

The (unbounded) 1D-wire medium consists of a set of infinitely long wires arranged in a square lattice with lattice constant $a$ (i.e. the spacing between the wires is $a$). The wires are assumed to be perfectly conducting, have radius $r_w$, and are oriented along the z-direction. The semi-infinite structure is obtained by replacing the region $z<0$ by air, as illustrated in Fig. 1. The wires are embedded in a dielectric with (relative) permittivity $\varepsilon_h$. In the following, the reflection problem in the homogenized structure is studied, and it is explained why an ABC is necessary.

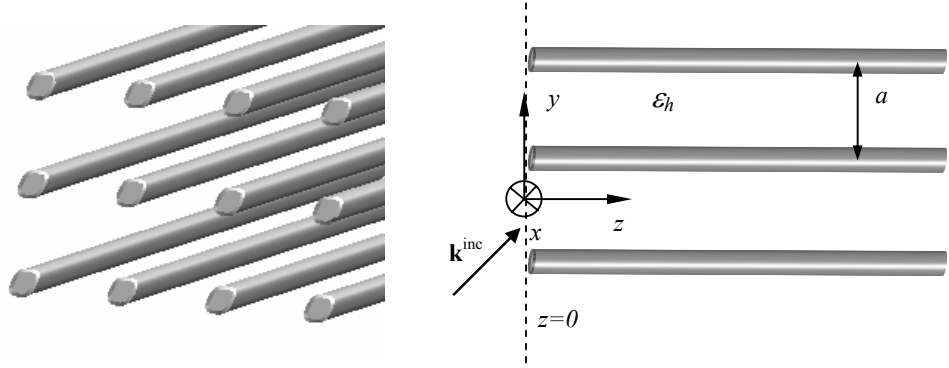

**Fig. 1** The truncated wire medium. The structure is periodic in the *xoy* plane.

### A. The homogenized wire medium

To begin with, the key properties of the homogenized wire medium are quickly reviewed. In [8] it was proved that the wire medium is an electromagnetic crystal with permittivity (relative to the host medium $\varepsilon_h$) $\varepsilon_{xx} = \varepsilon_{yy} = 1$ and

$$\varepsilon_{eff,zz} = 1 - \frac{\beta_p^2}{\beta_h^2 - k_z^2} \tag{1}$$

where $\beta_h = \beta\sqrt{\varepsilon_h}$, $\beta = \omega/c$ is the wave number in free-space, $\mathbf{k} = (k_x, k_y, k_z)$ is the wave vector, and $\beta_p$ is the plasma wave number. The permittivity dyadic predicts that there are 3 different plane wave solutions: a TEM mode, a TM mode and a TE mode. The dispersion characteristic of the modes is:

$$\beta_h = |k_z| \quad \text{(TEM mode)} \tag{2a}$$
$$\beta_h^2 = k^2 \quad \text{(TE mode)} \tag{2b}$$
$$\beta_h^2 = \beta_p^2 + k^2 \quad \text{(TM mode)} \tag{2c}$$

The polarization of the associated plane waves is of the form:

$$\mathbf{E} \propto \frac{\mathbf{k}_\|}{|\mathbf{k}_\||} e^{-j\mathbf{k}\cdot\mathbf{r}} \quad \text{(TEM mode)} \tag{3a}$$

$$\mathbf{E} \propto \frac{\mathbf{k}_\| \times \hat{\mathbf{u}}_z}{|\mathbf{k}_\| \times \hat{\mathbf{u}}_z|} e^{-j\mathbf{k}\cdot\mathbf{r}} \quad \text{(TE mode)} \tag{3b}$$

$$\mathbf{E} \propto \left( \frac{\mathbf{k}_{\parallel}}{\beta_h} + \frac{\beta_h^2 - k^2}{\beta_h^2 \varepsilon_{\mathit{eff},zz} - k^2} \frac{k_z}{\beta_h} \hat{\mathbf{u}}_z \right) e^{-j\mathbf{k}\cdot\mathbf{r}} \qquad \text{(TM mode)} \qquad (3c)$$

and $\mathbf{k}_{\parallel} = (k_x, k_y, 0)$. The corresponding magnetic field is given by:

$$\mathbf{H} = \frac{1}{\eta_h} \frac{\mathbf{k}}{\beta_h} \times \mathbf{E} \qquad (4)$$

where $\eta_h$ is the impedance of the host medium. It is important to note that in "conventional media" there are only two different plane wave solutions for a fixed frequency. The homogenized wire medium has not this property because it is characterized by strong spatial dispersion in the long wavelength limit [8].

### B. The scattering problem

The existence of three different electromagnetic modes has an important consequence: the usual boundary conditions (continuity of the tangential component of the electric and magnetic fields) at a planar interface between the (homogenized) wire medium and another dielectric medium are not enough to solve unambiguously a scattering problem. In fact, the corresponding system has one degree of freedom (because of the extra electromagnetic mode). To remove the indetermination an additional boundary condition is necessary.

To illustrate this difficulty let us consider the problem of plane wave incidence in a semi-infinite homogenized wire medium. For simplicity, we assume that the component of the incident wave vector parallel to the interface is $\mathbf{k}_{\parallel} = (0, k_y, 0)$, as depicted in Fig. 1. We consider that incident magnetic field is parallel to the interface (TM-z polarization) (for TE-z polarization the problem is trivial because the corresponding plane wave does not interact with the wires). Thus, the magnetic field is oriented along the $x$-direction, and the incident field can only excite the TEM and TM modes inside the wire medium. Therefore we can write that,

$$H_x = \left( e^{-\gamma_0 z} + \rho_e e^{+\gamma_0 z} \right) e^{-jk_y y} \qquad \text{(air side: } z < 0\text{)} \qquad (5a)$$

$$H_x = \left( B_1 e^{-j\beta_h z} + B_2 e^{-\gamma_{TM} z} \right) e^{-jk_y y} \qquad \text{(wire medium side: } z > 0\text{)} \qquad (5b)$$

where $\gamma_0 = \sqrt{k_y^2 - \beta^2}$, $\gamma_{TM} = \sqrt{\beta_p^2 + k_y^2 - \beta_h^2}$, $\rho_e$ is the reflection coefficient, and $B_1$ and $B_2$ are the amplitudes of the TEM and TM modes inside the wire medium. It can be easily verified that the continuity of the tangential component of the electric field is equivalent to the continuity of $\varepsilon_h^{-1} dH_x/dz$, where $\varepsilon_h$ is equal to unity at the air side. So, the classical boundary conditions impose that $H_x$ and $\varepsilon_h^{-1} dH_x/dz$ are continuous at the interface. Since the number of unknowns is three, the system is obviously underdetermined and the necessity for the ABC is evident.

Now the objective is to find an ABC that is able to accurately describe the electrodynamics of the wire medium. Which physical attribute of the wire medium can be used to formulate an ABC? There is one important property that hopefully may be useful: the electric current at the wire extremities must vanish at the interface. How can we use this property to devise an ABC?

From the results of [10], it can be verified that in the wire medium the homogenized fields satisfy (the averaging is in the *xoy* plane) the following (exact) relation:

$$\omega\varepsilon_0\varepsilon_h \mathbf{E}_{av}(z;\mathbf{k}_\parallel) = -\left(\mathbf{k}_\parallel + j\frac{d}{dz}\hat{\mathbf{u}}_z\right) \times \mathbf{H}_{av}(z;\mathbf{k}_\parallel) + \frac{j}{a^2}\int_{\partial A} \mathbf{J}_c(\mathbf{r})e^{j\mathbf{k}_\parallel \cdot \mathbf{r}}dl \qquad (6)$$

where $\mathbf{J}_c$ is the density of current at a generic wire with cross-section $\partial A$, and $dl$ is the arc length. The formula is also valid at the air side, provided we take $\mathbf{J}_c = 0$ and $\varepsilon_h = 1$. Now, we note that if we impose that $\mathbf{J}_c = 0$ at the wire ends the second term in the right-hand side vanishes at the interface (and in the particular is continuous). On the other hand, the tangential magnetic field is continuous at the interface and consequently the $z$-component of the first term in the right-hand side is also continuous. Thus, we conclude that z-component of the vector in the left-hand side is also continuous, i.e. the electromagnetic fields must satisfy the *additional boundary condition*:

$$\varepsilon_h E_{n,av} \quad \text{is continuous at the interface} \qquad (7)$$

where $E_{n,av}$ is the normal component (along $z$ for our geometry) of the homogenized electric field. We remind that by convention $\varepsilon_h = 1$ at the air side. Note that the ABC was derived using only the fact that $\mathbf{J}_c = 0$ at the wire ends. Note that the ABC is also consistent with the well-known fact that in the non-homogenized problem $\varepsilon_h E_n$ is continuous at the interface, and thus the same relation must hold for the homogenized field (given the geometry of this particular problem). In [14] a similar ABC was proposed to take into account the effect of diffusion in a semiconductor.

In the particular case in which the host medium is the same at both sides of the interface (e.g. the wires stand in air), (7) establishes that the normal component of the average fields must be continuous at the interface. The proposed solution is in a certain sense counterintuitive, because apparently it violates the rule that the normal component of the electric displacement vector, $\mathbf{D}$, in the homogenized structure is continuous (because the effective permittivity of the homogenized wire medium is certainly different from $\varepsilon_h$).

However that is not the case. To understand the reason it is important to remember that when spatial dispersion occurs, the permittivity dyadic has meaning only in the wave vector $\mathbf{k}$-space (which is the Fourier Transform dual of the $\mathbf{r}$-space). Of course, the permittivity dyadic can still be used to calculate the plane wave solutions in the homogenized structure (which describe the properties of the (spatial average) electromagnetic modes in the non-homogenized structure). The boundary conditions at an interface have meaning only in $\mathbf{r}$-space not in the $\mathbf{k}$-space. In general (at least for an unbounded spatially dispersive medium), the electric displacement vector $\mathbf{D}$ is related with the electric field by a spatial convolution, and not by a simple multiplication. When the field in the spatial domain consists of a sum of plane waves (associated with different wave vectors), $\mathbf{E}(\mathbf{r}) = \sum_i \mathbf{E}_i(\mathbf{r};\mathbf{k}_i)$, the corresponding electric displacement vector is given by $\mathbf{D}(\mathbf{r}) = \sum_i \overline{\overline{\varepsilon}}(\mathbf{k}_i) \cdot \mathbf{E}_i(\mathbf{r};\mathbf{k}_i)$. Therefore, there is neither contradiction nor redundancy between the continuity of the normal component of $\mathbf{D}$ and (7). It is also important to note that even though the z-component of the TEM mode vanishes, its contribution to $D_z$ is different from zero because the permittivity seen by this mode is infinite.

An important consequence of (7) is that when the wires stand in free-space not only the tangential components of the homogenized fields are continuous, but also the normal components. This is a very unusual and interesting phenomenon.

After simple calculations it is possible to verify that for the geometry under study, (7) is equivalent to impose that $\left[d^2 H_x/dz^2\right] = -\left(\beta_h^2 - \beta^2\right) H_x$ at the interface, where the rectangular brackets, $[\ ]$, represent the difference between a quantity calculated at the wire medium side and the same quantity calculated at the air side. Therefore, the boundary conditions necessary to solve the scattering problem can be summarized as:

$$\left[H_x\right] = 0; \quad \left[\varepsilon_h^{-1} dH_x/dz\right] = 0; \quad \left[d^2 H_x/dz^2\right] = -\left(\beta_h^2 - \beta^2\right) H_x \tag{8}$$

When the wires are embedded in air, $H_x$ and its first and second order derivatives are continuous at the interface. Using the boundary conditions in (5) we obtain the following linear system:

$$\begin{pmatrix} -1 & 1 & 1 \\ -\gamma_0 & -j\beta_h \varepsilon_h^{-1} & -\gamma_{TM} \varepsilon_h^{-1} \\ -\gamma_0^2 + \beta_h^2 - \beta^2 & -\beta_h^2 & \gamma_{TM}^2 \end{pmatrix} \begin{pmatrix} \rho_e \\ B_1 \\ B_2 \end{pmatrix} = \begin{pmatrix} 1 \\ -\gamma_0 \\ \gamma_0^2 - \beta_h^2 + \beta^2 \end{pmatrix} \tag{9}$$

Solving for $\rho_e$ we obtain:

$$\rho_e = -\frac{\gamma_0^2 - \varepsilon_h \gamma_0 \gamma_{TM} - j\beta_h \varepsilon_h \gamma_0 + j\beta_h \gamma_{TM} + \beta^2 - \beta_h^2}{\gamma_0^2 + \varepsilon_h \gamma_0 \gamma_{TM} + j\beta_h \varepsilon_h \gamma_0 + j\beta_h \gamma_{TM} + \beta^2 - \beta_h^2} \tag{10}$$

In case the wires are embedded in air the formula simplifies to:

$$\rho_e = -\frac{j\beta - \gamma_0}{j\beta + \gamma_0} \frac{\gamma_{TM} - \gamma_0}{\gamma_{TM} + \gamma_0} \tag{11}$$

This results give the reflection coefficient for a plane wave incident in the semi-infinite wire medium. It is important to note that (10) is very different from the result derived in [21] which applies when the wires are parallel to the interface. The reasons are that in the configuration studied in [21], there is no spatial dispersion, the TEM mode cannot be excited, and the wires can be treated as point dipoles. It can also be verified that the corresponding result derived in [17] is not compatible with (10).

In a subsequent section, we will prove that (11) is completely consistent with the exact analytical result for thin wires.

C. *Finite wire medium slab*

The proposed homogenization procedure allows us not only to study the propagation of waves in the semi-infinite wire medium, but also in a slab with wires of length $L$. The geometry of the problem is similar to that of Fig. 1, except that there is a second interface at $z = L$. In this case, assuming that the incident wave is as in section II.B, the magnetic field is given by:

$$H_x = \left(e^{-\gamma_0 z} + \rho_e e^{+\gamma_0 z}\right) e^{-jk_y y} \qquad \text{(air side: } z < 0\text{)} \tag{12a}$$

$$H_x = \left(B_1^+ e^{-j\beta_h z} + B_1^- e^{+j\beta_h z} + B_2^+ e^{-\gamma_{TM} z} + B_2^- e^{+\gamma_{TM} z}\right) e^{-jk_y y} \quad \text{(wire medium: } 0 < z < L\text{)} \tag{12b}$$

$$H_x = t_e e^{-\gamma_0 (z-L)} e^{-jk_y y} \qquad \text{(air side: } z > L\text{)} \tag{12c}$$

where $\rho_e$ and $t_e$ are the unknown reflection and transmission coefficients, respectively, and $B_1^\pm$ and $B_2^\pm$ are the amplitudes of the TEM and TM modes inside the wire medium slab.

Using the boundary conditions (8) at the interfaces $z=0$ and $z=L$ we readily obtain a 6×6 linear system can be solved numerically to find the unknowns.

## III. NUMERICAL VALIDATION

The objective of this section is to validate the new ABC. To this end, full wave numerical results are compared with the results predicted by homogenization theory.

Since it is not viable to analyze numerically a structure in which the wires are infinitely long, we consider in the simulations that are the wires are finite with length $L$. The geometry is described in section II.C. We numerically computed the reflection coefficient at the $z = 0$ interface using the periodic moment method (MoM). The radius of the wires is $r_w = 0.01a$, and thus $\beta_p a \approx 1.37$. The analytical results (i.e. the homogenization results) are obtained from (12) and matching the fields at the interfaces $z = 0$ and $z = L$.

In the first example, the length of the wires is $L = 2.0a$, and the incident angle is $\theta_{inc} = 45[\deg]$. The amplitude and the phase of the reflection coefficient are depicted in Fig. 2 and Fig. 3, respectively. The full line represents the numerical results computed with the MoM, and the dashed line represents the homogenization theory results. We also plotted (long dashed line) the results obtained when the ABC is not taken into account and only the TEM mode is considered inside the wire medium (note that the TM mode is cut-off for long wavelengths) (these results are computed by setting $B_2^+ = B_2^- = 0$ in (12) and imposing the continuity of the tangential electromagnetic fields at the interfaces). As seen, such approach completely fails to describe the reflection characteristic of the wire medium (even for very long wavelengths). This illustrates the importance of the new ABC. The agreement between the numerical results and the analytical results is good, especially for $\beta a \ll \pi$ (which defines together with $|\mathbf{k}_\parallel| a \ll \pi$ the range of application of the model (1)). Note that for small frequencies the reflection is negligible because most of the incoming energy is coupled to the TEM mode.

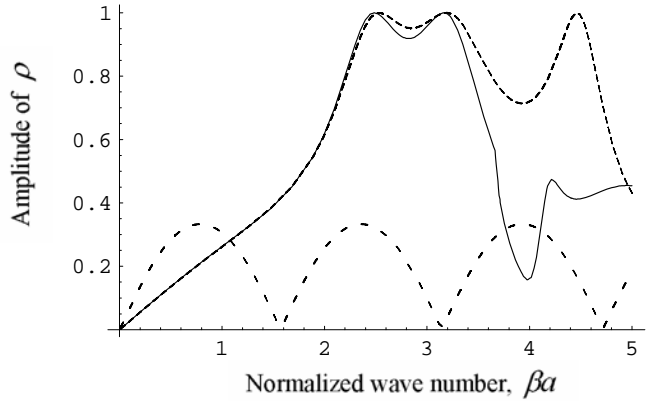

**Fig. 2** Amplitude of the reflection coefficient as a function of the normalized frequency (full line: numerical results; dashed line: homogenization theory; long dashed line: only TEM is considered).

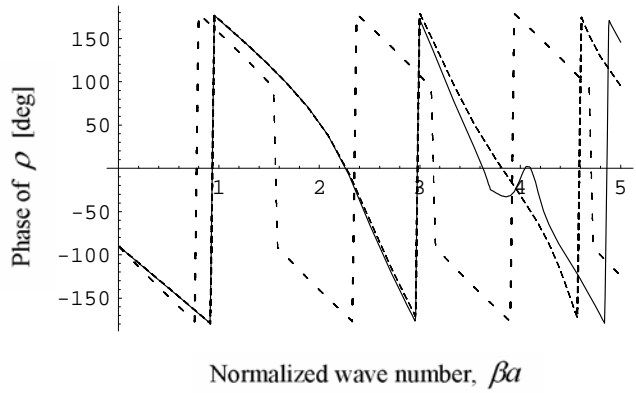

**Fig. 3** Phase of the reflection coefficient as a function of the normalized frequency (full line: numerical results; dashed line: homogenization theory; long dashed line: only TEM is considered).

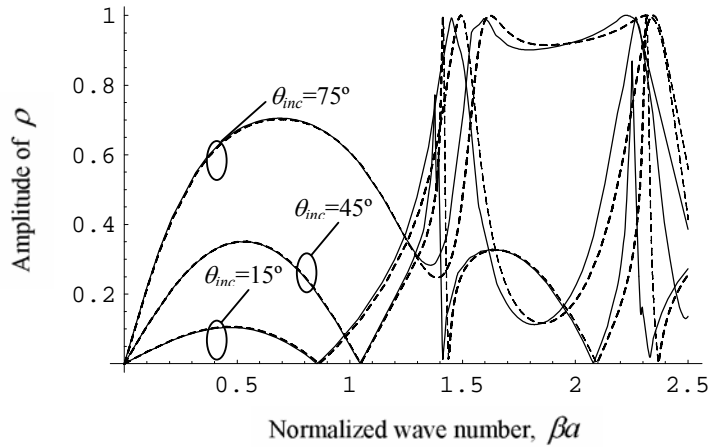

**Fig. 4** Amplitude of the reflection coefficient as a function of the normalized frequency (full line: numerical results; dashed line: homogenization theory).

The proposed homogenization model is also valid over a wide the range of incident angles and for the case in which the wires are embedded in a dielectric. This is illustrated in Fig. 4 for $\theta_{inc} = 15[\deg]$, $\theta_{inc} = 45[\deg]$, and $\theta_{inc} = 75[\deg]$ and $\varepsilon_h = 2.2$. As expected, the amplitude of the reflection coefficient decreases as the incident angle approaches the direction of normal incidence. The agreement between homogenization theory and the numerical results gets slightly worse when $\theta_{inc}$ increases. Similarly good agreement is obtained for the transmission coefficient.

Finally, in Fig. 5 the amplitude of the reflection coefficient is depicted for $\theta_{inc} = 45[\deg]$, $\varepsilon_h = 1.0$, and for the wire lengths: $L = 1.5a$, $L = 2.0a$ and $L = 3.0a$. As seen, the agreement is always good in the long wavelength limit.

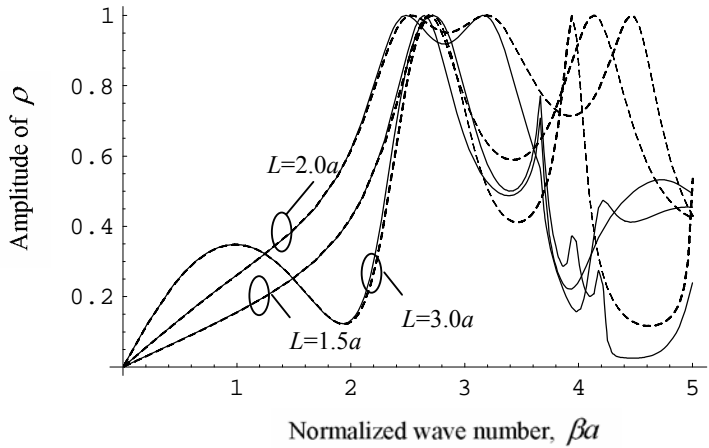

**Fig. 5** Amplitude of the reflection coefficient as a function of the normalized frequency (full line: numerical results; dashed line: homogenization theory).

## IV. SCATTERING BY A SEMI-INFINITE CRYSTAL

To further support the proposed ABC, in the second part of the paper we will calculate the exact analytical solution (within the thin-wire approximation) for the reflection coefficient of a plane wave that illuminates the semi-infinite wire medium, and prove that result is consistent with the ABC.

To begin with, in this section, we review the theory that is necessary to solve the scattering problem in a truncated periodic structure. In [20] it was proved that the reflection problem in a generic semi-infinite dielectric crystal can be reduced to an infinite linear system. The unknowns of the linear system are the amplitudes of the excited electromagnetic modes, and the coefficients of the corresponding infinite matrix only depend on the (generalized) electric dipole moment and on wave vector of the electromagnetic modes. We have extended these important results to the metallic case. To ease the readability of the paper the proof has been moved to Appendix A.

Consider that a 3D-periodic crystal is truncated at the plane $z=0$, being the region $z<0$ filled with air. The geometry of the resultant structure is as shown in Fig. 6. This semi-infinite crystal is invariant to translations along the primitive vectors $\mathbf{a}_1$ and $\mathbf{a}_2$. We assume that $\mathbf{a}_1$, $\mathbf{a}_2$ are parallel to the *xoy* plane. The area of the unit cell of the transverse lattice is $A_{cell} = |\mathbf{a}_1 \times \mathbf{a}_2|$.

The unbounded crystal is also invariant to translations along the primitive vector $\mathbf{a}_3$. We put $a_\perp = \mathbf{a}_3 \cdot \hat{\mathbf{u}}_z > 0$ (see Fig. 6). For future reference, we introduce the reciprocal lattice primitive vectors $\mathbf{b}_1$, $\mathbf{b}_2$ and $\mathbf{b}_3$, which are defined by the relations $\mathbf{a}_n \cdot \mathbf{b}_m = 2\pi \delta_{n,m}$, *n,m*=1,2,3.

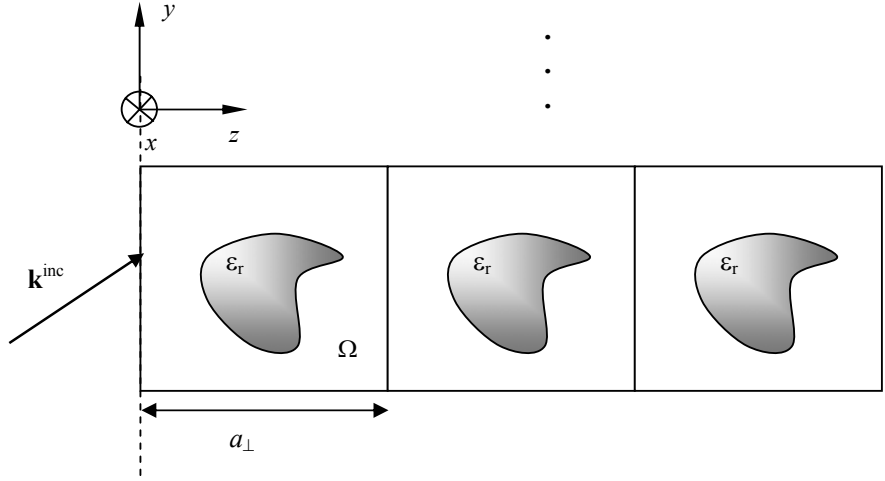

**Fig. 6** A plane wave impinges on a generic semi-infinite electromagnetic crystal.

The semi-infinite crystal is illuminated (from the air side) with a plane wave of the form $\mathbf{E}^{inc} = \mathbf{E}_0^{inc} \exp(-j\mathbf{k}^{inc}.\mathbf{r})$, where $\mathbf{k}^{inc} = \mathbf{k}_\| + k_\perp^{inc} \hat{\mathbf{u}}_z$ is the wave vector, and $\mathbf{E}_0^{inc}$ defines the polarization of the field. The wave vector satisfies the relation $\mathbf{k}^{inc}.\mathbf{k}^{inc} = \beta^2$, where $\beta = \omega/c$ is the free-space wave number, and $\omega$ is the angular frequency.

In the region $z > 0$ (the truncated crystal), the excited electric field can be expanded in terms of the electromagnetic Floquet modes $\mathbf{E}_n$, $n=1,2,..$, of the crystal:

$$\mathbf{E}(\mathbf{r}) = \sum_n c_n \mathbf{E}_n(\mathbf{r}; \mathbf{k}_n), \qquad z > 0 \tag{13}$$

where $c_n$ are the unknown coefficients of the expansion. The Floquet mode $\mathbf{E}_n$ is associated with the frequency $\omega$, and with the wave vector $\mathbf{k}_n$, i.e. $\mathbf{E}_n \exp(j\mathbf{k}_n.\mathbf{r})$ is periodic in the unbounded crystal. Since the component of the wave vector parallel to the interface must be preserved, the wave vector can be assumed of the form $\mathbf{k}_n = \mathbf{k}_\| + k_\perp^n \hat{\mathbf{u}}_z$. Also, because the electric field must satisfy a radiation condition, it follows that $\text{Re}(jk_\perp^n) > 0$ (assuming non-vanishing losses).

In Appendix A (see also [20]), it is proved that the unknown coefficients $c_n$ satisfy the following infinite linear system:

$$-\mathbf{E}_0^{inc} \delta_{\bar{\mathbf{J}},\mathbf{0}} = \frac{1}{2A_{cell}\gamma_{\bar{\mathbf{J}}}} \sum_n c_n \mathbf{k}_{\bar{\mathbf{J}}}^\pm \times \left( \mathbf{k}_{\bar{\mathbf{J}}}^\pm \times \frac{\mathbf{p}_n(\mathbf{k}_{\bar{\mathbf{J}}}^\pm)}{\varepsilon_0} \right) \frac{1}{e^{a_\perp(\gamma_{\bar{\mathbf{J}}} - jk_{\bar{\mathbf{J}},\perp}^n)} - 1}, \qquad \bar{\mathbf{J}} \text{ arbitrary} \tag{14}$$

where $\bar{\mathbf{J}} = (j_1, j_2)$ is a generic double-index of arbitrary integers, $\delta$ represents the Kronecker's $\delta$-symbol, and $\varepsilon_0$ is the permittivity of vacuum. We put,

$$\gamma_{\bar{\mathbf{J}}} = \sqrt{\mathbf{k}_{\bar{\mathbf{J}},\|}.\mathbf{k}_{\bar{\mathbf{J}},\|} - \beta^2}, \qquad \mathbf{k}_{\bar{\mathbf{J}},\|} = \mathbf{k}_\| + j_1 \mathbf{b}_{1,\|} + j_2 \mathbf{b}_{2,\|} \tag{15}$$

$$k_{\bar{\mathbf{J}},\perp}^n = k_\perp^n + j_1 \mathbf{b}_1.\hat{\mathbf{u}}_z + j_2 \mathbf{b}_2.\hat{\mathbf{u}}_z \tag{16}$$

$$\mathbf{k}_{\mathbf{J}}^{\pm} = \mathbf{k}_{\mathbf{\bar{J}},\parallel} \mp j\gamma_{\mathbf{\bar{J}}} \qquad (17)$$

where $\mathbf{b}_{1,\parallel}$ is the projection of $\mathbf{b}_1$ onto the *xoy* plane, etc. Also, the generalized electric dipole moment $\mathbf{p}_n$ was defined using:

$$\frac{\mathbf{p}_n(\mathbf{k}')}{\varepsilon_0} = \frac{1}{\beta^2} \int_{\partial D} (\hat{\mathbf{v}}' \times \nabla' \times \mathbf{E}_n) e^{+j\mathbf{k}'\cdot\mathbf{r}'} \, ds' + \int_{\Omega-D} (\varepsilon_r(\mathbf{r}') - 1) \mathbf{E}_n(\mathbf{r}') e^{+j\mathbf{k}'\cdot\mathbf{r}'} \, d^3\mathbf{r}' \qquad (18)$$

In the above, $\Omega = \{\alpha_1 \mathbf{a}_1 + \alpha_2 \mathbf{a}_2 + \alpha_3 \mathbf{a}_3 : |\alpha_1|, |\alpha_2| \leq 1/2 \text{ and } 0 \leq \alpha_3 \leq 1\}$ represents the unit cell of the unbounded crystal, $\varepsilon_r = \varepsilon_r(\mathbf{r})$ is the (periodic) relative permittivity of the structure, $D$ is the metallic region in the unit cell (if any), $\partial D$ represents the boundary surface of $D$, and $\hat{\mathbf{v}}$ is the outward unit vector normal to $\partial D$. Note that if $\mathbf{k}' = 0$ in (18), then $\mathbf{p}_n$ becomes the electric dipole moment (in a unit cell) for the Floquet mode $\mathbf{E}_n$.

On the other hand, in the air region $z < 0$ the scattered field is given by:

$$\mathbf{E}^s(\mathbf{r}) = \frac{-1}{2A_{cell}} \sum_{\mathbf{\bar{J}}} \frac{1}{\gamma_{\mathbf{\bar{J}}}} e^{-j\mathbf{k}_{\mathbf{\bar{J}}}^-\cdot\mathbf{r}} \sum_n c_n \mathbf{k}_{\mathbf{\bar{J}}}^- \times \left( \mathbf{k}_{\mathbf{\bar{J}}}^- \times \frac{\mathbf{p}_n(\mathbf{k}_{\mathbf{\bar{J}}}^-)}{\varepsilon_0} \right) \frac{1}{1 - e^{-a_\perp(\gamma_{\mathbf{\bar{J}}} + jk_{\mathbf{\bar{J}},\perp}^n)}}, \qquad z < 0 \quad (19)$$

where $\mathbf{k}_{\mathbf{\bar{J}}}^-$ is defined as in (17). So, assuming that $\text{Re}(\gamma_{\mathbf{\bar{J}}}) > 0$ for $\mathbf{\bar{J}} \neq \mathbf{0}$, i.e. that only one reflected mode propagates in free-space without attenuation (which is certainly true for relatively long wavelengths), the asymptotic expression of the scattered field as $z \to -\infty$ is:

$$\mathbf{E}^s(\mathbf{r}) \doteq \frac{-1}{2A_{cell}} \frac{1}{\gamma_0} e^{-j\mathbf{k}^{ref}\cdot\mathbf{r}} \sum_n c_n \mathbf{k}^{ref} \times \left( \mathbf{k}^{ref} \times \frac{\mathbf{p}_n(\mathbf{k}^{ref})}{\varepsilon_0} \right) \frac{1}{1 - e^{-a_\perp(\gamma_0 + jk_\perp^n)}}, \qquad z \to -\infty \quad (20)$$

where we put $\mathbf{k}^{ref} = \mathbf{k}_\parallel - k_\perp^{inc} \hat{\mathbf{u}}_z = \mathbf{k}_0^-$. In particular, the amplitude of the propagating (fundamental) reflected mode referred to the interface *z*=0, is:

$$\mathbf{E}_0^{ref} = \frac{-1}{2A_{cell}} \frac{1}{\gamma_0} \sum_n c_n \mathbf{k}^{ref} \times \left( \mathbf{k}^{ref} \times \frac{\mathbf{p}_n(\mathbf{k}^{ref})}{\varepsilon_0} \right) \frac{1}{1 - e^{-a_\perp(\gamma_0 + jk_\perp^n)}} \qquad (21)$$

The above results show that if somehow we are able to solve the linear system (14), then we can compute the solution of the electromagnetic problem in all space, and in particular the reflection coefficient at the interface.

## V. REFLECTION PROBLEM IN THE SEMI-INFINITE WIRE MEDIUM

Here, we use the formalism presented in the previous section to characterize the reflection of a plane wave by the wire medium. The geometry of problem is as in section II, except that we assume that the wires stand in air, i.e. $\varepsilon_h = 1$ (it can be verified that only this case has an analytical solution).

### A. Infinite linear system

In order to calculate the reflected field using (21), we need to solve the infinite linear system defined by (14). To this end, we characterize first the electromagnetic modes supported by the unbounded (non-truncated) wire medium. As is well known, the electromagnetic modes of this structure can be decomposed into 3 sets: transverse electromagnetic modes (TEM), transverse electric modes (TE), and the transverse magnetic modes (TM). The TEM modes are such that both $E_z$ and $H_z$ are zero, the TM modes are such that $H_z = 0$, and finally the TE modes are such that $E_z = 0$. It is noted that the dependency of the Floquet modes with the z variable is of the form $\exp(-jk_z z)$.

It is assumed that the thin-wire approximation, $r_w/a \ll 1$, holds. Under this assumption the density of current $\mathbf{J}_c$ (calculated at the surface of a generic wire) for an electromagnetic mode with wave vector $\mathbf{k}$, is of the form,

$$\mathbf{J}_c = \frac{I}{2\pi r_w} e^{-j\mathbf{k}\cdot\mathbf{r}} \hat{\mathbf{u}}_z \tag{22}$$

where $I$ is the current. Thus, apart from the propagation factor, it is assumed that the density of current is uniform and flows along the wire axis.

Within this approximation, it can be proved that the TE-Floquet modes degenerate into plane waves that propagate undisturbed in the wire medium (i.e. the TE-plane waves are not scattered by the wires).

In order to solve the infinite linear system (14), we calculate first the (generalized) dipole moment associated with the n-th Floquet mode with wave vector $\mathbf{k}_n$. Using (18) and the fact that the wires are embedded in air, it is found that:

$$\frac{\mathbf{p}_n(\mathbf{k}_\mathbf{J}^\pm)}{\varepsilon_0} = \frac{\eta_0}{j\beta} \int_{\partial D} \mathbf{J}_{c,n}(\mathbf{r}') e^{+j\mathbf{k}_\mathbf{J}^\pm \cdot \mathbf{r}'} ds' \tag{23}$$

where $\mathbf{J}_{c,n}$ is the density of current associated with the n-th mode, and the surface of the wire in the unit cell is defined by $\partial D = \{(x,y,z): 0 < z < a_\perp \text{ and } x^2 + y^2 = r_w^2\}$. Note that since the non-truncated structure is uniform in the z-direction the period $a_\perp$ can be chosen arbitrarily. Using (22) we obtain that:

$$\frac{\mathbf{p}_n(\mathbf{k}_\mathbf{J}^\pm)}{\varepsilon_0} = \hat{\mathbf{u}}_z \frac{\eta_0 I_n}{j\beta} \frac{J_0(|\mathbf{k}_{\mathbf{J},\|}^0| r_w)(e^{a_\perp(\pm\gamma_\mathbf{J} - jk_\perp^n)} - 1)}{(\pm\gamma_\mathbf{J} - jk_\perp^n)}, \qquad \mathbf{k}_{\mathbf{J},\|}^0 = \frac{2\pi}{a}(j_1, j_2, 0) \tag{24}$$

In the above, $I_n$ is the current along the wire in the unit cell (for the considered electromagnetic mode in the non-truncated structure), and $\eta_0$ is the free-space impedance.

We note that $I_n = 0$ for the TE-Floquet modes, and so the corresponding generalized dipole moment $\mathbf{p}_n(\mathbf{k}_\mathbf{J}^\pm)$ vanishes (within the thin-wire approximation). This proves that in (14) and (21) the parcels associated with the TE-modes also vanish, and therefore they can be discarded (i.e. the summation in the n variable can be restricted to the TEM and TM modes). This is not surprising, since as noted before the TE-modes are not scattered by the wires.

Substituting (24) into (14), taking into account that $k_{\mathbf{J},\perp}^n = k_\perp^n$ because the wires are arranged in a square lattice, and calculating the scalar product of both sides of the resulting equation with $\hat{\mathbf{u}}_z$, it is found that:

$$\mathbf{E}_0^{inc} \cdot \hat{\mathbf{u}}_z \delta_{\mathbf{J},0} = \frac{(\gamma_\mathbf{J}^2 + \beta^2) J_0(|\mathbf{k}_{\mathbf{J},\|}^0| r_w)}{2A_{cell} \gamma_\mathbf{J}} \sum_n c_n \frac{\eta_0 I_n}{j\beta} \frac{1}{\gamma_\mathbf{J} - jk_\perp^n}, \qquad \overline{\mathbf{J}} \text{ arbitrary} \tag{25}$$

Defining $A_n$ such that,

$$A_n = \frac{(\gamma_0^2 + \beta^2)}{2A_{cell} \gamma_0} \frac{\eta_0 I_n}{j\beta} c_n \tag{26}$$

and noting that the left-hand side of (25) vanishes for $\bar{\mathbf{J}} \neq 0$, we find that the unknowns $A_n$ satisfy:

$$\mathbf{E}_0^{inc}.\hat{\mathbf{u}}_z \delta_{\bar{\mathbf{J}},0} = \sum_n A_n \frac{1}{\gamma_{\bar{\mathbf{J}}} - jk_\perp^n}, \qquad \bar{\mathbf{J}} \text{ arbitrary} \tag{27}$$

Similarly, substituting (24) into (21) and remembering that $\mathbf{k}^{ref} = \mathbf{k}_0^-$, it is found that:

$$\mathbf{E}_0^{ref}.\hat{\mathbf{u}}_z = \sum_n A_n \frac{1}{\gamma_0 + jk_\perp^n} \tag{28}$$

Therefore, we can to compute the *z*-component of the reflected field using (28), by solving first the infinite linear system (27) with respect to the unknown variables $A_n$. As referred before, the summations in (27) and (28) can be restricted to the TM and TEM electromagnetic modes of the wire medium.

For a given $\mathbf{k}_\parallel$ (which only depends on the incident field), there is only one TEM mode that satisfies the correct radiation condition. Its propagation constant is known in closed-analytical form:

$$k_{\perp,TEM} = \beta \tag{29}$$

On the other hand, for a given $\mathbf{k}_\parallel$ there are infinitely many different (but countable) TM modes. Their propagation constants are of the form:

$$k_{\perp,TM}^n = -j\sqrt{\beta_{n,TM}^2(\mathbf{k}_\parallel) - \beta^2} \qquad n=1,2,3,\ldots \tag{30}$$

where $-\beta_{n,TM}^2(\mathbf{k}_\parallel)$ represent the eigenvalues of the (2D) Laplacian $\nabla^2 = \partial^2/\partial x^2 + \partial^2/\partial y^2$ associated with the Floquet eigenwaves that satisfy Dirichlet boundary conditions over the boundary of the metallic wires, and are independent of *z*. The eigenvalues $-\beta_{n,TM}^2(\mathbf{k}_\parallel)$ are real valued and unfortunately cannot be calculated in closed analytical form. However, they can be calculated numerically as explained for example in [25].

Next, it is convenient to define the following function in the complex plane (*w* is the complex variable):

$$f(w) = \sum_n A_n \frac{1}{w - jk_\perp^n} \tag{31}$$

We will suppose that the series is absolutely convergent in the complex plane (except at the poles $w = jk_\perp^n$). Equations (27) and (28) are equivalent to:

$$\mathbf{E}_0^{inc}.\hat{\mathbf{u}}_z \delta_{\bar{\mathbf{J}},0} = f(\gamma_{\bar{\mathbf{J}}}), \qquad \bar{\mathbf{J}} \text{ arbitrary} \tag{32}$$

$$\mathbf{E}_0^{ref}.\hat{\mathbf{u}}_z = -f(-\gamma_0) \tag{33}$$

In particular, putting $\bar{\mathbf{J}} = 0$ in (32) it is found that the reflection coefficient $\rho$ is given by:

$$\rho \equiv \frac{\mathbf{E}_0^{ref}.\hat{\mathbf{u}}_z}{\mathbf{E}_0^{inc}.\hat{\mathbf{u}}_z} = -\frac{f(-\gamma_0)}{f(\gamma_0)} \tag{34}$$

Therefore, provided we are able to calculate the unknown function $f$ then we can solve the problem analytically. This is discussed in the following sections. To begin with, in the next section the poles and zeros of $f$ are characterized.

B. Zeros and poles of function $f$

Equation (32) shows that function $f$ vanishes at $\gamma_{\bar{\mathbf{J}}}$ for $\bar{\mathbf{J}} \neq 0$. Thus, the zeros of $f$ are either purely imaginary numbers or purely real positive numbers. For convenience let $\left|\mathbf{k}_{\bar{\mathbf{J}}_m,\|}\right|$ ($m=1,2,\ldots$) represent the sequence of real numbers,

$$\left\{ \left|\mathbf{k}_{\bar{\mathbf{J}},\|}\right|^2 : \bar{\mathbf{J}} \neq 0 \right\} = \left\{ \left(k_x + \frac{2\pi}{a} j_1\right)^2 + \left(k_y + \frac{2\pi}{a} j_2\right)^2 : (j_1, j_2) \neq (0,0) \right\}$$

ordered in such a way that $\left|\mathbf{k}_{\bar{\mathbf{J}}_m,\|}\right| \leq \left|\mathbf{k}_{\bar{\mathbf{J}}_{m+1},\|}\right|$. Then $\breve{z}_m = \sqrt{\left|\mathbf{k}_{\bar{\mathbf{J}}_m,\|}\right|^2 - \beta^2}$ ($m=1,2,3\ldots$) are the zeros of function $f$, ordered in such a way that the real valued zeros satisfy $\breve{z}_m \leq \breve{z}_{m+1}$.

Even though it is not possible to write $\left|\mathbf{k}_{\bar{\mathbf{J}}_m,\|}\right|$ explicitly as a function of $m$, it is possible to write an asymptotic formula for $\breve{z}_m$. In fact, we have that:

$$m = \#\left\{ \bar{\mathbf{J}} : \left|\mathbf{k}_{\bar{\mathbf{J}},\|}\right| \leq \left|\mathbf{k}_{\bar{\mathbf{J}}_m,\|}\right| \text{ and } \bar{\mathbf{J}} \neq 0 \right\} \approx \#\left\{ \bar{\mathbf{J}} : \left|\mathbf{k}_{\bar{\mathbf{J}},\|}\right| \leq \breve{z}_m \right\} \approx \frac{\pi \breve{z}_m^2}{\left(\frac{2\pi}{a}\right)^2} \tag{35}$$

In the above, the symbol "#" represents the cardinal (number of elements) of a given set. The last identity results from the fact that in the wave vector space each index $\bar{\mathbf{J}}$ can be unambiguously assigned to a square with side $2\pi/a$, and from the fact that the area of the region $|\mathbf{k}| < \breve{z}_m$ is $\pi \breve{z}_m^2$. Using the previous result, we obtain the asymptotic formula:

$$\breve{z}_m \doteq \frac{2\pi}{a}\sqrt{\frac{m}{\pi}} \tag{36}$$

For future reference, we also note that:

$$\sum_{m=1}^{\infty} \left| \frac{1}{\breve{z}_{m+1}} - \frac{1}{\breve{z}_m} \right| < \infty \tag{37}$$

In fact, assuming that for $m > m_0$ the sequence $\breve{z}_m$ is real valued, we have that:

$$\sum_{m=m_0}^{\infty} \left| \frac{1}{\breve{z}_{m+1}} - \frac{1}{\breve{z}_m} \right| = \sum_{m=m_0}^{\infty} \left( \frac{1}{\breve{z}_m} - \frac{1}{\breve{z}_{m+1}} \right) = \lim_{m \to \infty} \left( \frac{1}{\breve{z}_{m_0}} - \frac{1}{\breve{z}_{m+1}} \right) = \frac{1}{\breve{z}_{m_0}} \tag{38}$$

The second identity is a consequence of $\breve{z}_m \leq \breve{z}_{m+1}$ when $\breve{z}_m$ is real valued.

On the other hand, (31) shows that the poles of $f$, denoted by $\breve{p}_n$, are given by:

$$\breve{p}_0 = jk_{\perp,TEM} = j\beta \quad ; \quad \breve{p}_n = jk_{\perp,TM}^n = \sqrt{\beta_{n,TM}^2(\mathbf{k}_\|) - \beta^2}, \quad n=1,2,\ldots \tag{39}$$

Note that except possibly for a finite number of purely imaginary poles, all the poles are real valued and positive. We assume that the sequence $\beta_{n,TM}^2(\mathbf{k}_\parallel)$ is increasing, so that the real poles satisfy $\breve{p}_n \leq \breve{p}_{n+1}$.

Next we prove that the poles and zeros of $f$ must alternate in the real line. Within the thin-wire approximation, it can be proven (see [7] for a related result) that $\lambda = \beta_{n,TM}^2$ ($n$=1,2,3…) satisfies the characteristic system:

$$0 = \sum_{\mathbf{J}} \frac{\left[J_0\left(\left\|\mathbf{k}_{\mathbf{J},\parallel}^0\right\|r_w\right)\right]^2}{\left|\mathbf{k}_{\mathbf{J},\parallel}\right|^2 - \lambda} \tag{40}$$

The above equation can be rewritten (rearranging the summation order) as:

$$0 = \sum_{m=0}^{\infty} \frac{d_m^2}{\left|\mathbf{k}_{\bar{\mathbf{J}}_m,\parallel}\right|^2 - \lambda} \tag{41}$$

where $d_m$ is some real constant and by definition $\mathbf{k}_{\bar{\mathbf{J}}_0,\parallel} \equiv (k_x, k_y, 0)$. The objective is to characterize the (real) roots ($\lambda = \beta_{n,TM}^2$) of the equation.

Let us assume first that $\lambda < \left|\mathbf{k}_{\bar{\mathbf{J}}_0,\parallel}\right|^2$. In that case it is obvious that the right-hand side of the equation is a positive quantity and so the equation has no solutions.

Consider now that $\left|\mathbf{k}_{\bar{\mathbf{J}}_{n-1},\parallel}\right|^2 < \lambda < \left|\mathbf{k}_{\bar{\mathbf{J}}_n,\parallel}\right|^2$, $n$=1,2,… (assuming for simplicity that we have a non-degenerate case: $\left|\mathbf{k}_{\bar{\mathbf{J}}_{n-1},\parallel}\right| \neq \left|\mathbf{k}_{\bar{\mathbf{J}}_n,\parallel}\right|$). It is convenient to rewrite the characteristic equation as:

$$\sum_{m=0}^{n-1} \frac{d_m^2}{\lambda - \left|\mathbf{k}_{\bar{\mathbf{J}}_m,\parallel}\right|^2} = \sum_{m=n}^{\infty} \frac{d_m^2}{\left|\mathbf{k}_{\bar{\mathbf{J}}_m,\parallel}\right|^2 - \lambda} \tag{42}$$

In the considered interval, the right-hand side of the equation is positive and grows monotonically to $+\infty$ as $\lambda$ approaches the upper extreme of the interval. On the other hand, the left-hand side of the equation is also positive and decays monotonically from $+\infty$ (at the lower extreme of the interval) to some constant (at the upper extreme) This proves that there exists exactly one solution of the characteristic equation in the interval $\left|\mathbf{k}_{\bar{\mathbf{J}}_{n-1},\parallel}\right|^2 < \lambda < \left|\mathbf{k}_{\bar{\mathbf{J}}_n,\parallel}\right|^2$. This solution is precisely $\lambda = \beta_{n,TM}^2$. Thus, we have proved that:

$$\left|\mathbf{k}_{\bar{\mathbf{J}}_{n-1},\parallel}\right|^2 \leq \beta_{n,TM}^2 \leq \left|\mathbf{k}_{\bar{\mathbf{J}}_n,\parallel}\right|^2 \qquad n=1,2,3,\ldots \tag{43}$$

This identity implies that for real valued poles/zeros we have that:

$$\breve{z}_{m-1} \leq \breve{p}_m \leq \breve{z}_m \qquad \text{(real valued poles/zeros)} \tag{44}$$

In the long wavelength limit, $\beta a \ll \pi$, the above identity is valid for arbitrary $m$=1,2,3…. For simplicity, we shall assume in the following sections that the zeros and the poles are real valued for $m \geq 1$. The pole $\breve{p}_0$ is always purely imaginary. Nonetheless the derived results will still be valid in the general case.

Because of (44) we have $\left|\dfrac{1}{\breve{p}_m} - \dfrac{1}{\breve{z}_m}\right| \leq \left|\dfrac{1}{\breve{z}_{m-1}} - \dfrac{1}{\breve{z}_m}\right|$ (for sufficiently large $m$). Therefore from (37) we find that:

$$\sum_{m=1}^{\infty} \left|\frac{1}{\breve{p}_m} - \frac{1}{\breve{z}_m}\right| < \infty \tag{45}$$

Also, for future reference, we note that (36) and (44) imply that:

$$\sum_{m=1}^{\infty} \frac{1}{|\breve{z}_m|^3} < +\infty \quad ; \quad \sum_{m=1}^{\infty} \frac{1}{|\breve{p}_m|^3} < +\infty \tag{46}$$

To conclude this section, we prove that $f$ given by (31) is uniformly bounded in sectors of the complex plane of the form $S(\theta_{\min}) = \{w : \theta_{\min} \leq |\theta| \leq \pi\}$ ($\theta$ is the polar angle) and a neighborhood of the pole $\breve{p}_0$. In fact,

$$|w - \breve{p}_n| \geq \breve{p}_n |\sin \theta_{\min}| \qquad n=1,2,\ldots \qquad w \in S(\theta_{\min}) \tag{47}$$

where the poles are assumed real, as referred before. So, from (31) we find that in the considered sector:

$$|f(w)| \leq \frac{1}{|\sin \theta_{\min}|} \sum_{n=1}^{\infty} \left|\frac{A_n}{\breve{p}_n}\right| + \left|\frac{A_0}{w - \breve{p}_0}\right| \tag{48}$$

The right-hand side is uniformly bounded outside a small neighborhood of $\breve{p}_0$. Note that the series in the right-hand side converges because we assume that the series in (31) is absolutely convergent (in particular it is absolutely convergent for $w = 0$).

## C. Construction of function f

Next, we will compute the unknown function $f$. The obvious idea is to write the function as an infinite product of monomials that depend on the zeros and poles, as suggested in [20] for a related problem. However, here things are not so simple.

In fact, in the problem under study the sequences of poles and zeros converge to infinity relatively slowly (see (46)), which prevents the convergence of some pertinent infinite products. Thus it is necessary to proceed with extra care (in [20] the situation is easier because the sequences of poles and zeros convergence to infinity extremely fast).

Another crucial issue concerns the uniqueness of $f$. In fact there are many different analytical functions with exactly the same poles and zeros. All these functions differ by a multiplication of a function of the form $\exp(g(w))$, where $g(w)$ is an entire function. So the important question is, "how to pick the correct function?" (this important issue was ignored in [20]). The answer is intrinsically related with the behavior of $f$ at infinity, as is shown in the proof presented next.

To begin with, we introduce the auxiliary functions $P_1$ and $P_2$ defined by (we assume without loss of generality that all the zeros and poles are different from zero):

$$P_1(w) = \prod_{n=0}^{\infty} \left(1 - \frac{w}{\breve{p}_n}\right) \exp\left(\frac{w}{\breve{p}_n} + \frac{1}{2}\left(\frac{w}{\breve{p}_n}\right)^2\right) \tag{49}$$

$$P_2(w) = \prod_{n=1}^{\infty}\left(1 - \frac{w}{\breve{z}_n}\right)\exp\left(\frac{w}{\breve{z}_n} + \frac{1}{2}\left(\frac{w}{\breve{z}_n}\right)^2\right) \tag{50}$$

The absolute and uniform convergence of the infinite products is guaranteed by (46) (see [26] pp. 56; the exponential factors are necessary to ensure convergence). Moreover, $P_1$ and $P_2$ are entire functions and the zeros of $P_1$ are precisely $w = \breve{p}_n$ ($n=0,1,2\ldots$), whereas the zeros of $P_2$ are $w = \breve{z}_n$ ($n=1,2\ldots$). The idea is to write $f$ in terms of $P_1$ and $P_2$.

Before that, it is important to estimate the order of growth of $P_1$ and $P_2$ when $|w| \to \infty$ (the reason will be clear ahead). An entire function $P$ has *order* (of growth) $\rho$ if for every $\varepsilon > 0$ the following relation holds,

$$\max|P(w)|_{|w|=r} \leq e^{r^{\rho+\varepsilon}} \tag{51}$$

for $r$ sufficiently large (see [26, pp. 63] for a more rigorous definition). Because of (36) and (44), it is possible to prove that the order of growth of $P_1$ and $P_2$ is $\rho = 2$ ([26, pp.69]).

Since the zeros of $P_1$ are exactly the same as the poles of function $f$, we conclude that $f P_1$ is an entire function that has exactly the same zeros as $f$. Moreover, it is possible to prove (see Appendix B) that the order of growth of $f P_1$ is also $\rho = 2$. Because of this we can apply Hadamard's factorization theorem [26, pp.74] that establishes the following important result,

$$f(w)P_1(w) = e^{g(w)}P_2(w) \tag{52}$$

where $g$ is some polynomial of degree no larger than $\rho = 2$. Rearranging the previous equation, it is found that:

$$f(w) = e^{g(w)}\frac{1}{\left(1 - \frac{w}{\breve{p}_0}\right)}\prod_{n=1}^{\infty}\frac{\left(1 - \frac{w}{\breve{z}_n}\right)\exp\left(\frac{w}{\breve{z}_n} + \frac{1}{2}\left(\frac{w}{\breve{z}_n}\right)^2\right)}{\left(1 - \frac{w}{\breve{p}_n}\right)\exp\left(\frac{w}{\breve{p}_n} + \frac{1}{2}\left(\frac{w}{\breve{p}_n}\right)^2\right)} \tag{53}$$

where $g$ still represents some (unknown) polynomial of degree no larger than 2. Now because of (45), we can apply a Lemma enunciated in Appendix B to conclude that,

$$Q(w) = \prod_{n=1}^{\infty}\frac{\left(1 - \frac{w}{\breve{z}_n}\right)}{\left(1 - \frac{w}{\breve{p}_n}\right)} \tag{54}$$

is absolutely and uniform convergent in compact sets of the complex plane that do not contain the poles. But this implies that we can rewrite (53) as follows:

$$f(w) = e^{\tilde{g}(w)}\frac{1}{\left(1 - \frac{w}{\breve{p}_0}\right)}\prod_{n=1}^{\infty}\frac{\left(1 - \frac{w}{\breve{z}_n}\right)}{\left(1 - \frac{w}{\breve{p}_n}\right)} \tag{55}$$

$$\tilde{g}(w) = g(w) + w \sum_{n=1}^{\infty} \left( \frac{1}{\tilde{z}_n} - \frac{1}{\breve{p}_n} \right) + \frac{w^2}{2} \sum_{n=1}^{\infty} \left( \frac{1}{\tilde{z}_n^2} - \frac{1}{\breve{p}_n^2} \right) \qquad (56)$$

Apart from the unknown function $\tilde{g}$, we have obtained a representation for $f$ in terms of canonical products/quotients of monomials that depend on its zeros/poles. From the above, it is clear $\tilde{g}$ is a polynomial of degree no larger than 2.

In Appendix B, it is proven that because the zeros and poles alternate in the real line (see (44)) the infinite product given by (54) has the following bounds ($\theta_{\min} > 0$ is arbitrary):

$$\frac{|\sin \theta_{\min}|}{\left|1 - \frac{w}{\breve{p}_1}\right|} \leq |Q(w)| \leq \frac{1}{|\sin \theta_{\min}|}, \qquad w \in S(\theta_{\min}) \qquad (57)$$

The above formula assumes that $\tilde{z}_m$ and $\breve{p}_m$ are positive and real valued for $m \geq 1$, but similar bounds can be obtained for another more general situation.

Because of the lower bound of $|Q|$ in the sector $S(\theta_{\min})$, we conclude from (55) that $f$ will grow exponentially to infinity in some regions of the complex plane, unless the unknown polynomial $\tilde{g}$ is a constant. However $f$ cannot grow to infinity exponentially as $w \to \infty$ because (48) established that $|f|$ is uniformly bounded in $S(\theta_{\min})$ (as $w \to \infty$). So, we conclude that $\tilde{g}$ is necessarily a constant. Therefore we have the result:

$$f(w) = \frac{C}{\left(1 - \frac{w}{\breve{p}_0}\right)} \prod_{n=1}^{\infty} \frac{\left(1 - \frac{w}{\tilde{z}_n}\right)}{\left(1 - \frac{w}{\breve{p}_n}\right)} \qquad (58)$$

where $C$ is some constant. It will be seen in the next section that the knowledge of $C$ is not relevant in the problem studied in this paper.

D. Reflection coefficient and comparison with the ABC solution

Substituting (58) in (34) we easily find that the reflection coefficient is:

$$\rho \equiv \frac{\mathbf{E}_0^{ref} \cdot \hat{\mathbf{u}}_z}{\mathbf{E}_0^{inc} \cdot \hat{\mathbf{u}}_z} = -\frac{\breve{p}_0 - \gamma_0}{\breve{p}_0 + \gamma_0} \prod_{n=1}^{\infty} \frac{\tilde{z}_n + \gamma_0}{\tilde{z}_n - \gamma_0} \frac{\breve{p}_n - \gamma_0}{\breve{p}_n + \gamma_0} \qquad (59)$$

In the long wavelength limit, the above formula can be rewritten as explained next. Noting that for $\beta a \ll \pi$ all the poles and zeros are real valued for $m \geq 1$, and that $\gamma_0$ is in the imaginary axis, we have:

$$\rho = -\frac{j\beta - \gamma_0}{j\beta + \gamma_0} \frac{\breve{p}_1 - \gamma_0}{\breve{p}_1 + \gamma_0} e^{2\gamma_0 \delta} \qquad (60)$$

$$\delta = \frac{1}{|\gamma_0|} \sum_{n=1}^{\infty} \left( \arctan\left(\frac{|\gamma_0|}{\tilde{z}_n}\right) - \arctan\left(\frac{|\gamma_0|}{\breve{p}_{n+1}}\right) \right) \approx \sum_{n=1}^{\infty} \left( \frac{1}{\tilde{z}_n} - \frac{1}{\breve{p}_{n+1}} \right) \qquad (61)$$

The second identity in (61) is valid if $\beta a \ll \pi$. The parameter $\delta$ has units of length and its physical meaning will be discussed ahead. Note that (44) implies that:

$$0 \leq \delta \leq \sum_{n=1}^{\infty}\left(\frac{1}{\tilde{z}_n} - \frac{1}{\tilde{z}_{n+1}}\right) = \frac{1}{\tilde{z}_1}, \qquad \beta a \ll \pi, \quad ka \ll \pi \qquad (62)$$

But since $\tilde{z}_1 \approx 2\pi/a$, we obtain that to a first approximation $\delta$ has the following bounds:

$$0 \leq \delta \leq \frac{a}{2\pi}, \qquad \beta a \ll \pi, \quad ka \ll \pi \qquad (63)$$

Next, the exact result (60) is compared with the approximate formula (11) obtained in the first part of the paper using the ABC. First of all, we note that the definition of the reflection coefficient is consistent in the two cases because the ratio between the reflected and incident magnetic fields is equal to the ratio between the z-component of the reflected electric field and the z-component of the incident field. Also note that (2c) is the dispersion characteristic of the fundamental TM mode in the wire medium, and so $\beta_{1,TM}^2(\mathbf{k}_\parallel) \approx \beta_p^2 + k_\parallel^2$. Thus, $\gamma_0$ and $\breve{p}_1 = jk_{\perp,TM}^1$ are defined consistently with the symbols $\gamma_0$ and $\gamma_{TM}$ of section II.B when $k_x = 0$ and $\varepsilon_h = 1$. Therefore we obtain that,

$$\rho = \rho_e e^{2\gamma_0 \delta} \qquad (64)$$

Thus, apart from the (complex) exponential factor (which as proved ahead is very close to unity), the reflection coefficient calculated with proposed ABC agrees exactly with the analytical formula (60). This demonstrates the generality and effectiveness of the ABC. In the next section, it is explained how the exponential factor can be incorporated in the model of homogenized the wire medium.

To conclude, we refer that the approach that was used to characterize the reflection coefficient in the semi-infinite wire medium can also be applied to other related structures (such as the semi-infinite parallel-plate medium), as briefly explained in Appendix C.

*E. Virtual interface*

Since $\gamma_0$ is the propagation constant in the z-direction (in the air region), it is clear from (64) and from elementary transmission line theory, that the semi-infinite wires can be replaced by their equivalent homogenized model without affecting the reflection coefficient, provided the interface with air is displaced from a distance $\delta$ from the actual physical interface. More specifically, we have the situation depicted in Fig. 7.

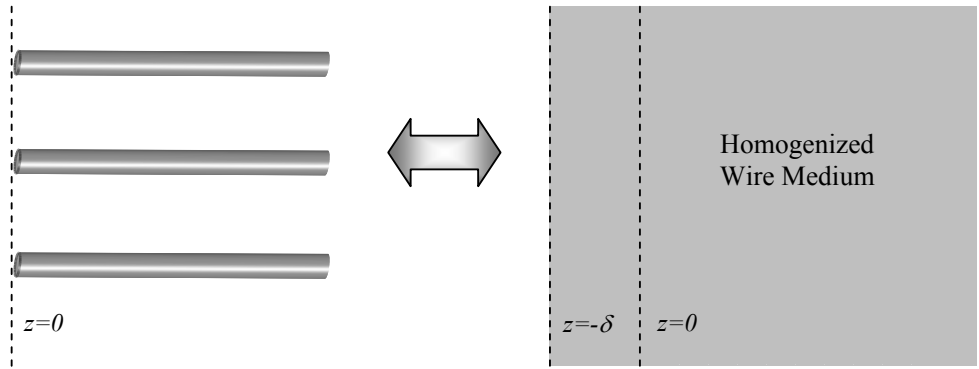

**Fig. 7** Definition of virtual interface.

This means that in the homogenized model the interface with air is not coincident with the physical interface. Due to this reason we say that $z = -\delta$ is a virtual interface. This concept was also used in other works [23]-[24]. Note that in general $\delta = \delta(\beta, \mathbf{k}_\parallel)$, i.e. $\delta$ depends on the frequency and on the direction of the incoming wave. However, in the long wavelength limit and to a first approximation it is possible to write $\delta \approx \delta(\beta = 0, \mathbf{k}_\parallel = 0)$. From (63) it is known that for long wavelengths, and within the thin wire approximation, $0 \leq \delta \leq 0.16a$.

Obviously, $\delta$ is also a function of the radius $r_w$ of the wires. Using (61) we numerically calculated $\delta$ as a function of $r_w$ in the limit $\beta = 0$ and $\mathbf{k}_\parallel = 0$. To this end, the poles $\tilde{p}_n$ were numerically calculated using the method proposed in [25] and the thin-wire approximation. As shown in Fig. 8, $\delta$ increases with $r_w$.

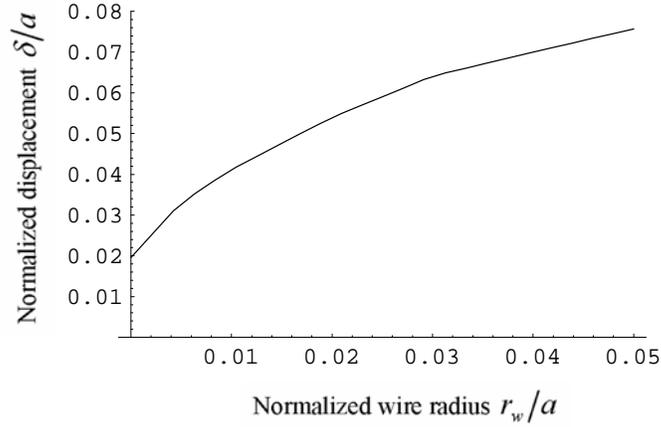

**Fig. 8** Normalized displacement $\delta/a$ as a function of the radius of the normalized wire radius $r_w/a$

We numerically verified that even though considering the effect of virtual interfaces helped improving the agreement between homogenization theory and the full wave results, the difference was not very relevant. In fact, for thin wires (i.e. within the scope of application of our theory) the effect of virtual interfaces seems to be negligible and so the ABC is sufficient to accurately homogenize the structure.

## VI. CONCLUSIONS

In this paper, we proposed a new boundary condition to homogenize the wire medium. The proposed ABC imposes that the normal component of the average electric field multiplied by the host permittivity is continuous across the interface. It was shown that this ABC does not violate the continuity of the normal component of the electric displacement vector, which is still observed. The proposed theory was successfully tested against full wave numerical simulations, being the agreement good even for relatively small wavelengths, thin slabs, and also in the case in which the wires are embedded in a dielectric. It was shown that if the ABC was not considered and single mode propagation was assumed instead, then homogenization theory would completely fail to describe the scattering problem.

To demonstrate the generality and efficacy of the ABC, we analytically solved the problem of scattering of plane wave by an array of semi-infinite thin metallic wires. To this end, we reduced the electromagnetic problem to a linear system of infinite dimension. We showed that the solution the infinite linear system is intrinsically related with the problem of calculating an analytical function from the knowledge of its zeros, poles, and behavior at infinity. Using Hadamard's factorization theorem and other results, we were able to rigorously reconstruct the required function, and in this way compute the reflection coefficient at the interface between the wire medium and air. It was shown that the exact reflection coefficient was compatible with effective medium theory, provided a virtual interface was defined and the ABC was considered.

The new ABC paves the way to the homogenization of some important structures such as the bed of nails substrate (ground plane with metallic pins) used to synthesize artificial magnetic conductors. These results will be reported in a future communication.

## APPENDIX A

Here, we prove the results enunciated in section IV. The results are an extension to the metallic case of the formalism described in [18]-[20]. The proof presented [20] is based on the local field theory, while our equivalent approach is based on integral representation theory.

### A. The unbounded crystal

To begin with, we characterize the electromagnetic Floquet modes in the unbounded periodic crystal. A generic Floquet mode $\mathbf{E}$ associated with the free-space wave number $\beta = \omega/c$ and the wave vector $\mathbf{k}$ has the following integral representation (the identity is valid in the distributional sense and not as principal value integral):

$$\mathbf{E}(\mathbf{r}) = \int_{\partial D} \overline{\overline{\mathbf{G}}}_p(\mathbf{r}|\mathbf{r}';\mathbf{k}) \cdot (\hat{\mathbf{v}}' \times \nabla' \times \mathbf{E}) ds' + \int_{\Omega-D} \overline{\overline{\mathbf{G}}}_p(\mathbf{r}|\mathbf{r}';\mathbf{k}) \cdot \beta^2 (\varepsilon_r(\mathbf{r}') - 1) \mathbf{E}(\mathbf{r}') d^3\mathbf{r}' \tag{A1}$$

where all the symbols are defined as in section II, and the Green dyadic $\overline{\overline{\mathbf{G}}}_p = \overline{\overline{\mathbf{G}}}_p(\mathbf{r}|\mathbf{r}';\mathbf{k})$ is defined by,

$$\overline{\overline{\mathbf{G}}}_p = \left( \overline{\overline{\mathbf{I}}} + \frac{1}{\beta^2} \nabla \nabla \right) \Phi_p \tag{A2}$$

where $\overline{\overline{\mathbf{I}}}$ is the identity dyadic, and $\Phi_p$ is the lattice Green function that satisfies:

$$\nabla^2 \Phi_p + \beta^2 \Phi_p = -\sum_{\mathbf{I}} \delta(\mathbf{r} - \mathbf{r}' - \mathbf{r}_\mathbf{I}) e^{-j\mathbf{k} \cdot (\mathbf{r} - \mathbf{r}')} \tag{A3}$$

In the above, $\delta$ represents Dirac's distribution, $\mathbf{r}_\mathbf{I} = i_1 \mathbf{a}_1 + i_2 \mathbf{a}_2 + i_3 \mathbf{a}_3$ is a lattice point, and $\mathbf{I} = (i_1, i_2, i_3)$ is a multi-index of arbitrary integers. In [22] several closed-form representations of the Green function are derived. Of particular interest for this work, is the spectral-like representation of the Green function, which establishes that for $|z - z'| < a_\perp$:

$$\Phi_p = \sum_{\bar{\mathbf{J}}} \frac{1}{A_{cell}} \frac{1}{2\gamma_{\bar{\mathbf{J}}}} \left( e^{-\gamma_{\bar{\mathbf{J}}} |z-z'|} + \sum_{\pm} \frac{e^{\pm \gamma_{\bar{\mathbf{J}}}(z-z')}}{e^{a_\perp (\gamma_{\bar{\mathbf{J}}} \pm jk_{\bar{\mathbf{J}},\perp})} - 1} \right) e^{-j\mathbf{k}_{\bar{\mathbf{J}},\|} \cdot (\mathbf{r} - \mathbf{r}')} \tag{A4}$$

where $\gamma_{\bar{\mathbf{J}}}$ is given by (15), $\mathbf{k}_{\bar{\mathbf{J}}} = \mathbf{k} + j_1 \mathbf{b}_1 + j_2 \mathbf{b}_2$, $\mathbf{k}_{\bar{\mathbf{J}},\|}$ is the projection of $\mathbf{k}_{\bar{\mathbf{J}}}$ onto the *xoy* plane, $k_{\bar{\mathbf{J}},\perp} = \mathbf{k}_{\bar{\mathbf{J}}} \cdot \hat{\mathbf{u}}_z$, and $\bar{\mathbf{J}} = (j_1, j_2)$ is a double-index of arbitrary integers.

### B. The semi-infinite crystal

Consider now the semi-infinite crystal depicted in Fig. 6. As referred in section II, the semi-infinite crystal is illuminated with the plane wave $\mathbf{E}^{inc} = \mathbf{E}_0^{inc} \exp(-j\mathbf{k}^{inc} \cdot \mathbf{r})$. The total electric field can be written as $\mathbf{E} = \mathbf{E}^{inc} + \mathbf{E}^s$, where $\mathbf{E}^s$ is the scattered field.

Using standard Green function arguments it can be proved that the scattered field has the integral representation:

$$\mathbf{E}^s(\mathbf{r}) = \int_{\bigcup_l \partial D_l} \overline{\overline{\mathbf{G}}}_L(\mathbf{r}|\mathbf{r}';\mathbf{k}_\|) \cdot (\hat{\mathbf{v}}' \times \nabla' \times \mathbf{E}) ds' + \int_{\bigcup_l \Omega_l - D_l} \overline{\overline{\mathbf{G}}}_L(\mathbf{r}|\mathbf{r}';\mathbf{k}_\|) \beta^2 (\varepsilon_r(\mathbf{r}') - 1) \mathbf{E}(\mathbf{r}') d^3\mathbf{r}' \quad (A5)$$

where $\Omega_l = \Omega + l\mathbf{a}_3$ ($l=0,1,2,\ldots$) is the translation of the unit cell along the vector $l\mathbf{a}_3$, $D_l = D + l\mathbf{a}_3$ is the translation of the metallic region $D$ along the vector $l\mathbf{a}_3$, and $\partial D_l$ represents the boundary surface of $D_l$. The Green dyadic $\overline{\overline{\mathbf{G}}}_L = \overline{\overline{\mathbf{G}}}_L(\mathbf{r}|\mathbf{r}';\mathbf{k}_\|)$ is the solution of:

$$\nabla \times \nabla \times \overline{\overline{\mathbf{G}}}_L = \beta^2 \overline{\overline{\mathbf{G}}}_L + \overline{\overline{\mathbf{I}}} \sum_{\overline{\mathbf{I}}} \delta(\mathbf{r} - \mathbf{r}' - \mathbf{r}_{\overline{\mathbf{I}}}) e^{-j\mathbf{k}_\| \cdot (\mathbf{r} - \mathbf{r}')} \quad (A6)$$

where $\mathbf{r}_{\overline{\mathbf{I}}} = i_1 \mathbf{a}_1 + i_2 \mathbf{a}_2$, and $\overline{\mathbf{I}} = (i_1, i_2)$ is a multi-index of arbitrary integers. The Green dyadic is given by,

$$\overline{\overline{\mathbf{G}}}_L = \left(\overline{\overline{\mathbf{I}}} + \frac{1}{\beta^2} \nabla\nabla\right) \Phi_L \quad (A7a)$$

$$\Phi_L(\mathbf{r}|\mathbf{r}';\mathbf{k}_\|) = \frac{1}{2 A_{cell}} \sum_{\overline{\mathbf{J}}} \frac{1}{\gamma_{\overline{\mathbf{J}}}} e^{-\gamma_{\overline{\mathbf{J}}} |z - z'|} e^{-j\mathbf{k}_{\overline{\mathbf{J}},\|} \cdot (\mathbf{r} - \mathbf{r}')} \quad (A7b)$$

where $\mathbf{k}_{\overline{\mathbf{J}},\|}$ is defined as in (15).

Equation (A5) can be rewritten as:

$$\mathbf{E}^s(\mathbf{r} + m\mathbf{a}_3) = \left(\overline{\overline{\mathbf{I}}} + \frac{1}{\beta^2} \nabla\nabla\right) \cdot \sum_{l=0}^{\infty} \left( \int_{\partial D} \Phi_L(\mathbf{r} + m\mathbf{a}_3 | \mathbf{r}' + l\mathbf{a}_3; \mathbf{k}_\|)(\hat{\mathbf{v}}' \times \nabla' \times \mathbf{E})\big|_{\mathbf{r}' + l\mathbf{a}_3} ds' \right.$$

$$\left. + \beta^2 \int_{\Omega - D} \Phi_L(\mathbf{r} + m\mathbf{a}_3 | \mathbf{r}' + l\mathbf{a}_3; \mathbf{k}_\|)(\varepsilon_r(\mathbf{r}') - 1) \mathbf{E}(\mathbf{r}' + l\mathbf{a}_3) d^3\mathbf{r}' \right) \quad (A8)$$

On the other hand, for $z > 0$ the total field can be expanded into electromagnetic Floquet modes, as established by (13). In order to calculate the unknown coefficients $c_n$ we substitute (13) in (A8) and use the fact that $\mathbf{E}_n$ is a Floquet mode. It is found that:

$$\mathbf{E}^s(\mathbf{r}) = \left(\overline{\overline{\mathbf{I}}} + \frac{1}{\beta^2} \nabla\nabla\right) \cdot \sum_n c_n \left( \int_{\partial D} \Phi_p^{unil}(\mathbf{r}|\mathbf{r}';\mathbf{k}_n)(\hat{\mathbf{v}}' \times \nabla' \times \mathbf{E}_n) ds' \right.$$

$$\left. + \beta^2 \int_{\Omega - D} \Phi_p^{unil}(\mathbf{r}|\mathbf{r}';\mathbf{k}_n)(\varepsilon_r(\mathbf{r}') - 1) \mathbf{E}_n(\mathbf{r}') d^3\mathbf{r}' \right) \quad (A9)$$

where,

$$\Phi_p^{unil}(\mathbf{r}|\mathbf{r}';\mathbf{k}_n) = \sum_{l=0}^{\infty} \Phi_L(\mathbf{r}|\mathbf{r}' + l\mathbf{a}_3; \mathbf{k}_\|) e^{-j\mathbf{k}_n \cdot l\mathbf{a}_3}$$

$$= \frac{1}{2 A_{cell}} \sum_{\overline{\mathbf{J}}} \frac{1}{\gamma_{\overline{\mathbf{J}}}} e^{-j\mathbf{k}_{\overline{\mathbf{J}},\|} \cdot (\mathbf{r} - \mathbf{r}')} \left( e^{-\gamma_{\overline{\mathbf{J}}} |z - z'|} + \frac{e^{+\gamma_{\overline{\mathbf{J}}} (z - z')}}{e^{a_\perp (\gamma_{\overline{\mathbf{J}}} + jk_{\overline{\mathbf{J}},\perp}^n)} - 1} \right) \quad (A10)$$

The last identity is valid for $|z-z'|<a_\perp$, and was obtained using (A7b) and noting that the resultant series in the $l$ index is a geometrical series. We put $k_{\mathbf{J},\perp}^n = (\mathbf{k}_n + j_1\mathbf{b}_1 + j_2\mathbf{b}_2)\cdot\hat{\mathbf{u}}_z$, consistently with (16). But using (A4), $\Phi_p^{unil}$ can be rewritten as:

$$\Phi_p^{unil}(\mathbf{r}|\mathbf{r}';\mathbf{k}_n) = \Phi_p(\mathbf{r}|\mathbf{r}';\mathbf{k}_n) - \frac{1}{2A_{cell}}\sum_{\mathbf{J}}\frac{1}{\gamma_{\mathbf{J}}}e^{-j\mathbf{k}_{\mathbf{J},\parallel}\cdot(\mathbf{r}-\mathbf{r}')}\frac{e^{-\gamma_{\mathbf{J}}(z-z')}}{e^{a_\perp(\gamma_{\mathbf{J}}-jk_{\mathbf{J},\perp}^n)}-1} \quad (A11)$$

Substituting the previous result in (A9), using (A1) and (13), it is found for $\mathbf{r} \in \Omega$ that:

$$\mathbf{E}^s(\mathbf{r}) - \mathbf{E}(\mathbf{r}) = \frac{-1}{2A_{cell}}\sum_{\mathbf{J}}\frac{1}{\gamma_{\mathbf{J}}}e^{-j\mathbf{k}_{\mathbf{J},\parallel}\cdot\mathbf{r}}e^{-\gamma_{\mathbf{J}}z}\left(\beta^2\overline{\overline{\mathbf{I}}} - \mathbf{k}_{\mathbf{J}}^\pm\mathbf{k}_{\mathbf{J}}^\pm\right)\sum_n c_n \frac{\mathbf{p}_n(\mathbf{k}_{\mathbf{J}}^+)}{\varepsilon_0}\frac{1}{e^{a_\perp(\gamma_{\mathbf{J}}-jk_{\mathbf{J},\perp}^n)}-1} \quad (A12)$$

where $\mathbf{k}_{\mathbf{J}}^\pm$ is defined by (17), and $\mathbf{p}_n$ is defined by (18).

Since $\mathbf{E} = \mathbf{E}^{inc} + \mathbf{E}^s$ and because $\mathbf{k}_{\mathbf{J}}^\pm\cdot\mathbf{k}_{\mathbf{J}}^\pm = \beta^2$, we finally obtain that:

$$-\mathbf{E}^{inc}(\mathbf{r}) = \frac{1}{2A_{cell}}\sum_{\mathbf{J}}\frac{1}{\gamma_{\mathbf{J}}}e^{-j\mathbf{k}_{\mathbf{J},\parallel}\cdot\mathbf{r}}e^{-\gamma_{\mathbf{J}}z}\sum_n c_n\, \mathbf{k}_{\mathbf{J}}^\pm \times \left(\mathbf{k}_{\mathbf{J}}^\pm \times \frac{\mathbf{p}_n(\mathbf{k}_{\mathbf{J}}^+)}{\varepsilon_0}\right)\frac{1}{e^{a_\perp(\gamma_{\mathbf{J}}-jk_{\mathbf{J},\perp}^n)}-1}, \quad \mathbf{r}\in\Omega \quad (A13)$$

But the above identity obviously implies that the unknown coefficients $c_n$ satisfy the infinite linear system (14), as we wanted to prove. On the other hand, using (A9) and (A10) we readily obtain (19).

## APPENDIX B

In this Appendix, we describe some auxiliary results related with the convergence and asymptotic behavior of some infinite products.

*Order of growth of $f P_1$*

Here we prove that the order of growth of $f P_1$ is the same as the order of growth of $P_1$. Using (31) and noting that $P_1(\breve{p}_n) = 0$, it is found that:

$$|f(w)P_1(w)|_{|w|=r} = \left|\sum_n \frac{A_n}{w-\breve{p}_n}P_1(w)\right|$$
$$\leq \sum_{|\breve{p}_n|>2r}\left|\frac{A_n}{w-\breve{p}_n}\right||P_1(w)| + \sum_{|\breve{p}_n|\leq 2r}\left|A_n \frac{P_1(w)-P_1(\breve{p}_n)}{w-\breve{p}_n}\right| \quad (B1)$$

On the other hand, we have that:

$$\left|\frac{1}{w-\breve{p}_n}\right| \leq \frac{2}{|\breve{p}_n|}, \qquad \text{for } |\breve{p}_n| > 2r = 2|w| \quad (B2)$$

and,

$$\left|\frac{P_1(w)-P_1(\breve{p}_n)}{w-\breve{p}_n}\right| \leq \max_{|u|\leq 2r}|P_1'(u)|, \qquad \text{for } |\breve{p}_n| \leq 2r = 2|w| \quad (B3)$$

Substituting the previous results in (B1), we obtain after further manipulations that:

$$|f(w)P_1(w)|_{|w|=r} \leq 2\left(\sum_{n=0}^{\infty}\left|\frac{A_n}{\breve{p}_n}\right|\right)\left(\max_{|u|=r}|P_1(u)|+r\max_{|u|\leq 2r}|P_1'(u)|\right) \tag{B4}$$

But since the order of growth of $P_1'$ is the same as the order of growth of $P_1$ [26, pp. 67], it is also evident that the order of $fP_1$ is the same as the order of $P_1$.

*Convergence of Q*

**Lemma**: Suppose that $z_n$ and $p_n$ are sequences of complex numbers such that $|z_n|,|p_n|\to\infty$ and $\sum_n\left|\frac{1}{p_n}-\frac{1}{z_n}\right|<\infty$. Then the infinite product $Q(w)$ given by (54) converges uniformly in compact sets of the complex plane that do not contain elements of the $p_n$ sequence.

In fact, it is known that if $g_n(w)$ is a sequence of complex functions such that $|g_n(w)|\leq\alpha_n$ in some domain $D$ of the complex plane, with $\sum_n\alpha_n<\infty$, then the infinite product $\prod_n(1+g_n(w))$ converges uniformly in $D$ [26]. The proof of the Lemma follows from the enunciated result by putting $g_n(w)=\frac{w}{1-w/p_n}\left(\frac{1}{p_n}-\frac{1}{z_n}\right)$.

*Bounds for Q*

In this section it is proven that the function $Q$ given by (54) satisfies (57). For simplicity, we assume that $\breve{z}_m$ and $\breve{p}_m$ are positive and real valued for $m\geq 1$. We remind (see section III.B) that $\breve{z}_m$ and $\breve{p}_m$ are increasing sequences.

To begin with, we note that if $a$ and $b$ are positive numbers then,

$$\left|1-\frac{w}{a}\right|\leq\left|1-\frac{w}{b}\right| \Leftrightarrow \begin{cases} |w|\leq\dfrac{2\cos\theta}{\dfrac{1}{a}+\dfrac{1}{b}} & \text{if } a\leq b \\ |w|\geq\dfrac{2\cos\theta}{\dfrac{1}{a}+\dfrac{1}{b}} & \text{if } a\geq b \end{cases} \tag{B5}$$

where $\theta$ is the polar angle. In particular, replacing $a$ for $\breve{z}_m$ and $b$ for $\breve{p}_m$, and noting that because of (44) $\breve{p}_m\leq\breve{z}_m$, it is evident that if $\operatorname{Re}\{w\}\leq 0$ (i.e. $\cos\theta\leq 0$) then $|Q(w)|\leq 1$.

Suppose now that $\operatorname{Re}\{w\}>0$ and define $N=N(w)$ such that:

$$|w|<\frac{2\cos\theta}{\dfrac{1}{\breve{z}_{N+1}}+\dfrac{1}{\breve{p}_{N+1}}} \quad\text{and}\quad |w|\geq\frac{2\cos\theta}{\dfrac{1}{\breve{z}_N}+\dfrac{1}{\breve{p}_N}} \tag{B6}$$

The above definition is possible because $1/\breve{z}_m+1/\breve{p}_m$ can always be assumed strictly decreasing. Then, using (B5), it is clear that

$$|Q(w)| \le \prod_{n=N+1}^{\infty} \frac{\left|1-\frac{w}{\breve{z}_n}\right|}{\left|1-\frac{w}{\breve{p}_n}\right|} = \prod_{n=N+1}^{\infty} \frac{\left|1-\frac{w}{\breve{z}_n}\right| \left|1-\frac{w}{\breve{p}_{n+1}}\right|}{\left|1-\frac{w}{\breve{p}_{n+1}}\right| \left|1-\frac{w}{\breve{p}_n}\right|} \qquad (B7)$$

Noting that:

$$\prod_{n=N+1}^{\infty} \frac{\left|1-\frac{w}{\breve{p}_{n+1}}\right|}{\left|1-\frac{w}{\breve{p}_n}\right|} = \left|\prod_{n=N+1}^{\infty} \frac{1-\frac{w}{\breve{p}_{n+1}}}{1-\frac{w}{\breve{p}_n}}\right| = \frac{1}{\left|1-\frac{w}{\breve{p}_{N+1}}\right|} \qquad (B8)$$

we find that:

$$|Q(w)| \le \frac{1}{\left|1-\frac{w}{\breve{p}_{N+1}}\right|} \prod_{n=N+1}^{\infty} \frac{\left|1-\frac{w}{\breve{z}_n}\right|}{\left|1-\frac{w}{\breve{p}_{n+1}}\right|} \qquad (B9)$$

But because of (44) we have that $\breve{z}_n \le \breve{p}_{n+1}$. Using (B6), it is also simple to verify that:

$$\frac{2\cos\theta}{\frac{1}{\breve{z}_n}+\frac{1}{\breve{p}_{n+1}}} \ge \frac{2\cos\theta}{\frac{1}{\breve{z}_{N+1}}+\frac{1}{\breve{p}_{N+1}}} > |w| \qquad , n \ge N+1 \qquad (B10)$$

Thus, using again (B5), we obtain:

$$|Q(w)| < \frac{1}{\left|1-\frac{w}{\breve{p}_{N+1}}\right|} \qquad (B11)$$

But since we assume that $\breve{p}_m$ is positive and real, it is easy to prove that:

$$|Q(w)| \le \frac{1}{|\sin\theta_{\min}|} \qquad \text{for } w \in S(\theta_{\min}) \qquad (B12)$$

This proves that $|Q|$ has an upper bound in the considered sector. The lower bound for $|Q|$ can be easily derived by applying similar arguments to the function $1/Q$.

## APPENDIX C

In this Appendix, we prove that the formalism used to solve the reflection problem in the truncated wire medium can also be applied to the semi-infinite parallel-plate medium. The parallel-plate medium consists of an infinite set of parallel metallic plates (see [27]). Here, we suppose that the plates are normal to the *x*-direction, and that the distance between the plates is *a*. The interface with air is at *z*=0. Let us consider the case in which the incoming wave has the magnetic field along the *y*-direction, and the component of the incident wave vector parallel to the interface is $\mathbf{k}_\| = (k_x, 0, 0)$. The objective is to compute the reflection coefficient at the interface. This problem was solved almost sixty years ago using the Wiener-Hopf method [28]-[29]. The solution of the problem can also be found in [27].

Using the formalism developed in this paper it is straightforward to verify that the reflection coefficient is still given by (59), but now the poles and the zeros are given by,

$$\breve{p}_n = \sqrt{\left(\frac{\pi}{a}n\right)^2 - \beta^2} \qquad n=0,1,2,\ldots \tag{C1}$$

$$\breve{z}_{2n} = \gamma_n \quad ; \quad \breve{z}_{2n-1} = \gamma_{-n} \quad ; \quad n=1,2,\ldots \tag{C2}$$

where $\gamma_n = \sqrt{\left(\frac{2\pi}{a}n + k_x\right)^2 - \beta^2}$, assuming without loss of generality that $k_x \geq 0$ (so that $\breve{z}_n$ is an increasing sequence). Our result and the formula presented in [27] seem remarkably different (apart from the differences in the notation). In fact, it can be verified that in both formulas the operator $\prod$ acts exactly over the same parcels. However the formula presented in [27] differs from ours by the term $-\exp(-2\ln 2\, \gamma_0 a/\pi)$ (in our notation). The lack of "–" sign in our formula is easy to justify: in fact in [27] the definition of $\rho$ also differs from a "–" sign from our definition. But what about the term $\exp(-2\ln 2\, \gamma_0 a/\pi)$? Is our formula wrong? The answer is no: it can be verified that both formulas yield exactly the same result when they are numerically evaluated. So, why has the formula in [27] the extra term $\exp(-2\ln 2\, \gamma_0 a/\pi)$? The reason is that although in both formulas $\prod$ acts over the same parcels, the sequence of multiplications is completely different in the two cases. The situation is the same as in non-absolutely convergent series: if the order of summation of an infinite number of terms is rearranged then the result of the sum may be different. This result helps understanding why in section III we had to proceed so carefully to construct the unknown function $f$.

## ACKNOWLEDGEMENT

This work was funded by Fundação para Ciência e a Tecnologia under project POSI 34860/99.## REFERENCES

[1] W. Rotman, "Plasma simulation by artificial dielectrics and parallel-plate media", IRE Trans. Antennas Propag. 10, pp. 82-95, Jan. 1962.
[2] D. R. Smith, W.J. Padilla, D.C. Vier, S.C. Nemat-Nasser, S. Schultz, "Composite Medium with Simultaneously Negative Permeability and Permittivity", Phys. Rev. Letts. 84, pp.4184-4187, May 2000.
[3] R. Shelby, D. R. Smith and S. Schultz, "Experimental verification of a negative index of refraction", *Science*, 292, 77 (2001).
[4] Stefan Enoch, Gérard Tayeb, et al. "A Metamaterial for Directive Emission", Phys. Rev. Letts., 89, 213902(1-4), 2002.
[5] G. Shvets, "Photonic approach to making a material with negative index of refraction" Phys Rev B, 67, 035109(1-8) (2003).
[6] A. Alù and N. Engheta, "Achieving transparency with plasmonic and metamaterial coatings" accepted to appear in Phys Rev E.
[7] M. Silveirinha, C. A. Fernandes, "Homogenization of 3D- Connected and Non-Connected Wire Metamaterials", IEEE Trans. on Microwave Theory and Tech., Vol. 53 , No. 4 , pp. 1418 - 1430, April, 2005, Special Issue on Metamaterialsat the interface. This problem was solved almost sixty years ago using the Wiener-Hopf method [28]-[29]. The solution of the problem can also be found in [27].

Using the formalism developed in this paper it is straightforward to verify that the reflection coefficient is still given by (59), but now the poles and the zeros are given by,

$$\breve{p}_n = \sqrt{\left(\frac{\pi}{a}n\right)^2 - \beta^2} \qquad n=0,1,2,\ldots \tag{C1}$$

$$\breve{z}_{2n} = \gamma_n \quad ; \quad \breve{z}_{2n-1} = \gamma_{-n} \quad ; \quad n=1,2,\ldots \tag{C2}$$

where $\gamma_n = \sqrt{\left(\frac{2\pi}{a}n + k_x\right)^2 - \beta^2}$, assuming without loss of generality that $k_x \geq 0$ (so that $\breve{z}_n$ is an increasing sequence). Our result and the formula presented in [27] seem remarkably different (apart from the differences in the notation). In fact, it can be verified that in both formulas the operator $\prod$ acts exactly over the same parcels. However the formula presented in [27] differs from ours by the term $-\exp(-2\ln 2\, \gamma_0 a/\pi)$ (in our notation). The lack of "–" sign in our formula is easy to justify: in fact in [27] the definition of $\rho$ also differs from a "–" sign from our definition. But what about the term $\exp(-2\ln 2\, \gamma_0 a/\pi)$? Is our formula wrong? The answer is no: it can be verified that both formulas yield exactly the same result when they are numerically evaluated. So, why has the formula in [27] the extra term $\exp(-2\ln 2\, \gamma_0 a/\pi)$? The reason is that although in both formulas $\prod$ acts over the same parcels, the sequence of multiplications is completely different in the two cases. The situation is the same as in non-absolutely convergent series: if the order of summation of an infinite number of terms is rearranged then the result of the sum may be different. This result helps understanding why in section III we had to proceed so carefully to construct the unknown function $f$.

## ACKNOWLEDGEMENT

This work was funded by Fundação para Ciência e a Tecnologia under project POSI 34860/99.

## REFERENCES

[1] W. Rotman, "Plasma simulation by artificial dielectrics and parallel-plate media", IRE Trans. Antennas Propag. 10, pp. 82-95, Jan. 1962.
[2] D. R. Smith, W.J. Padilla, D.C. Vier, S.C. Nemat-Nasser, S. Schultz, "Composite Medium with Simultaneously Negative Permeability and Permittivity", Phys. Rev. Letts. 84, pp.4184-4187, May 2000.
[3] R. Shelby, D. R. Smith and S. Schultz, "Experimental verification of a negative index of refraction", *Science*, 292, 77 (2001).
[4] Stefan Enoch, Gérard Tayeb, et al. "A Metamaterial for Directive Emission", Phys. Rev. Letts., 89, 213902(1-4), 2002.
[5] G. Shvets, "Photonic approach to making a material with negative index of refraction" Phys Rev B, 67, 035109(1-8) (2003).
[6] A. Alù and N. Engheta, "Achieving transparency with plasmonic and metamaterial coatings" accepted to appear in Phys Rev E.
[7] M. Silveirinha, C. A. Fernandes, "Homogenization of 3D- Connected and Non-Connected Wire Metamaterials", IEEE Trans. on Microwave Theory and Tech., Vol. 53 , No. 4 , pp. 1418 - 1430, April, 2005, Special Issue on Metamaterials

[8] P.A. Belov, R. Marqués, et al., "Strong spatial dispersion in wire media in the very large wavelength limit", Physical Review B 67, 113103 (2003).
[9] L. Landau, E. Lifchitz, "Electrodynamics of continuous media", Theoretical Physics, vol. 8.
[10] M. Silveirinha, C. A. Fernandes, "Homogenization of Metamaterial Surfaces and Slabs: The Crossed Wire Mesh Canonical Problem", IEEE Trans. on Antennas and Propagation, Vol. 53 , No. 1 , pp. 59-69 , Jan., 2005, Special Issue on Artificial Magnetic Conductors, Soft/Hard Surfaces, and other Complex Surfaces.
[11] V. Agranovich and V.Ginzburg, "Spatial dispersion in crystal optics and the theory of excitons" (Wiley- Interscience, NY, 1966)
[12] J. L. Birman and J. J.Sein, Phys. Rev. B 6, 2482 (1972)
[13] G. Agarwal, D. Pattanayak, and E.Wolf, Phys. Rev. B 10, 1447 (1974)
[14] W. A. Davis, C. M. Krowne, "The Effects of Drift and Diffusion in Semiconductors on Plane Wave Interaction at Interfaces", IEEE Trans. on AP, vol. 36, nº1, Jan. 1998, pp. 97-103.
[15] S.I. Pekar, Zh. Eksp. Teor. Fiz. 33, 1022 (1957). [Soviet Phys. JETP 6, 785 (1958)].
[16] K. Henneberger, Phys. Rev. Lett. 80, 2889 (1998).
[17] I.S. Nefedov, A.J. Viitanen, "Effects of spatial dispersion in wire medium, formed by two mutually orthogonal wire lattices", Proc. of Bianisotropics'2004, Ghent, Belgium, pp. 86-89 (I.S. Nefedov, A.J. Viitanen, and S.A. Tretyakov, "Electromagnetic wave refraction at an interface of a double wire medium", http://arxiv.org/abs/cond-mat/0509610)
[18] G. Mahan and G. Obermair, "Polaritons at surfaces" Phys. Rev. 183, 834 (1969)
[19] C. A. Meade, "Formally closed solution for a crystal with spatial dispersion", Phys. Rev. B 17, 4644-4651, 1978.
[20] P. Belov, C. Simovski, "Boundary conditions for interfaces of electromagnetic crystals and the generalized Ewald-Oseen extinction principle", Phys. Rev. B 73, 045102-(1-14) (2006).
[21] P. Belov, S. Tretyakov, and A. Viitanen, *J. Electromagn. Waves Appl.* **16**, 1153-1165, 2002.
[22] M. Silveirinha, C. A. Fernandes "A New Acceleration Technique with Exponential Convergence Rate to Evaluate the Periodic Green Function", IEEE Trans. on Antennas and Propagation, Vol. 53 , No. 1 , pp. 347 - 355 , Jan., 2005.
[23] C.R. Simovski, S.A. Tretyakov, A. H. Sihvola, M. Popov, *European Physical Journal: Applied Physics*, vol. 9, no. 3, pp. 233-240, (2000).
[24] C.R. Simovski, B. Sauviac, *European Physical Journal: Applied Physics*, vol. 17, no. 1, pp. 11-20 (2002).
[25] M. Silveirinha, C. A. Fernandes, "Efficient Calculation of the Band Structure of Artificial Materials With Cylindrical Metallic Inclusions", IEEE Trans. on Microwave Theory and Tech. , Vol. 51 , No. 5 , pp. 1460 - 1466 , May, 2003.
[26] Robert M. Young, "An Introduction to Nonharmonic Fourier Series", Academic Press, 1980
[27] Robert E. Collin, "Field Theory of Guided Waves", 2$^{nd}$ Ed. IEEE Press, 1991, pp. 664-671.
[28] J. F. Carlson and A. E. Heins, "The reflection of electromagnetic waves by an infinite set of plates", *Quart. Appl. Math.*, vol. 4, pp. 313-329, Jan. 1947
[29] J. F. Carlson and A. E. Heins, "The reflection of electromagnetic waves by an infinite set of plates – part II", *Quart. Appl. Math.*, vol. 5, pp. 82-88, Apr. 1947